\documentclass[useAMS,usenatbib]{mn2e}
\usepackage{amssymb}
\usepackage{amsmath}
\usepackage{graphicx}
\usepackage{BibDef}
\usepackage{xcolor}
\usepackage{hyperref}
\hypersetup{
  colorlinks=true,        % false: boxed links; true: coloured links
  linkcolor=blue,         % colour of internal links
  citecolor=blue,         % colour of links to bibliography
}

\usepackage{ulem}
\def\lsim{\mathrel{\rlap{\lower 3pt \hbox{$\sim$}} \raise 2.0pt \hbox{$<$}}}
\def\gsim{\mathrel{\rlap{\lower 3pt \hbox{$\sim$}} \raise 2.0pt \hbox{$>$}}}
\def\msun{\rm {M_{\large \odot}}}

\title[Post-Newtonian MBH Dynamics] {Post-Newtonian evolution of massive black hole triplets in galactic nuclei -- II. Survey of the parameter space}
\author[Bonetti et al.]{Matteo Bonetti$^{1,2}$, Francesco Haardt$^{1,2}$, Alberto Sesana$^3$ \& Enrico Barausse$^{4,5}$\\
$^1$DiSAT, Universit\`a degli Studi dell'Insubria, Via Valleggio 11, 22100 Como, Italy\\
$^2$INFN, Sezione di Milano-Bicocca, Piazza della Scienza 3, 20126 Milano, Italy\\
$^3$Institute of Gravitational Wave Astronomy and School of Physics and Astronomy, University of
Birmingham, Edgbaston, Birmingham \\ \, B15 2TT, United Kingdom\\
$^4$CNRS, UMR 7095, Institut d'Astrophysique de Paris, 98 bis Bd Arago, 75014 Paris, France\\
$^5$Sorbonne Universit\'es, UPMC Univesit\'e Paris 6, UMR 7095, Institut d'Astrophysique de Paris, 98 bis Bd Arago, 75014 Paris, France
}

\begin{document}

\date{~}

\pagerange{\pageref{firstpage}--\pageref{lastpage}} \pubyear{2017}

\maketitle

\label{firstpage}

\begin{abstract} 
Massive black hole binaries (MBHBs) are expected to form at the centre of merging galaxies during the hierarchical assembly of the cosmic structure, and are expected to be the loudest sources of gravitational waves (GWs) in the low frequency domain. However, because of the dearth of energy exchanges with background stars and gas, many of these MBHBs may stall at separations too large for GW emission to drive them to coalescence in less than a Hubble time. Triple MBH systems are then bound to form after a further galaxy merger, triggering a complex and rich dynamics that can eventually lead to MBH coalescence. Here we report on the results of a large set of numerical simulations, where MBH triplets are set in spherical stellar potentials and MBH dynamics is followed through 2.5 post-Newtonian order in the equations of motion. From our full suite of simulated systems we find that a fraction $\simeq 20-30$ \% of the MBH binaries that would otherwise stall are led to coalesce within a Hubble time. The corresponding coalescence timescale peaks around 300 Myr, while the eccentricity close to the plunge, albeit small, is non-negligible ($\lsim 0.1$). We construct and discuss marginalised probability distributions of the main parameters involved and, in a companion paper of the series, we will use the results presented here to forecast the contribution of MBH triplets to the GW signal in the nHz regime probed by Pulsar Timing Array experiments. 
\end{abstract}


%%%%%%%%%%%%%%%%%%%%%%%%%%%%%%%%%%%%%%%%%%%%%%%%%%%%%%%%%%%%%%%%%%%%%%%%%%%%%%%%%%%%
%%%%%%%%%%%%%%%%%%%%%%%%%%%%%%%%%%%%%%%%%%%%%%%%%%%%%%%%%%%%%%%%%%%%%%%%%%%%%%%%%%%%
\section{Introduction}
\label{sec:Intro}

Massive black holes (MBHs) are ubiquitous in the nuclei of nearby spheroids \citep[see][and references therein]{Kormendy2013}, and are recognised to be a fundamental ingredient in the process of galaxy formation and evolution. Indeed, the tight correlations existing among the mass of the central MBH and the properties of the host galaxy \citep[see, e.g.,][]{Ferrarese2000,Gebhardt2000,Tremaine2002,Ferrarese2004,Ferrarese2006} indicate that galaxies and MBHs follow a linked evolutionary path during the formation history of cosmic structures. 
It is therefore understood that MBHs were commonly residing at the centres of galaxies at all cosmic epochs. This very circumstance, when framed in the bottom-up hierarchical clustering of cold dark matter overdensities, leads to the inevitable conclusion that a large number of MBH binaries (MBHBs) formed during the build-up of the large scale structure \citep{Begelman1980}. 

MBHBs are expected to be the loudest sources of gravitational radiation in the nHz-mHz frequency range \citep{Haehnelt1994,Jaffe2003,Wyithe2003,Enoki2004,Sesana2004,Sesana2005,Jenet2005,Rhook2005,Barausse2012,Klein2016}, a regime partially covered by the LISA interferometer \citep{LISA2013,LISA2017}, and by existing 
Pulsar Timing Array (PTA) experiments \citep{Desvignes2016,NANOGrav2015,Reardon2016,Verbiest2016}.
The observability of MBHBs by { LISA} and PTAs relies on the ability of the two black holes to coalesce within a Hubble time after a galaxy merger\footnote{See however \citet{Dvorkin2017} for the stochastic gravitational wave (GW) background under the hypothesis that all MBHBs stalled. That background, while suppressed relative to the case of efficient MBH mergers, would still be potentially observable by PTAs in the SKA era.}. 

The evolution of a MBHB in a galactic potential can be divided in three different stages \citep{Begelman1980}. Initially, driven by dynamical friction against stars and gas, the two MBHs migrate 
toward the centre of the newly formed spheroid to form a bound binary system. The subsequent evolution of the binary depends on the properties of the surrounding environment in the nucleus of the merger galaxy remnant. In gas rich galaxies (wet mergers), further shrinking can be caused by the interaction with a massive circumbinary disc or with incoherent pockets of accreted gas clouds \citep{Dotti2007,Cuadra2009,Nixon2011,Goicovic2017}. However, most of the simulations exploring these scenarios are highly idealised, usually lacking realistic prescriptions for cooling, fragmentation, star formation and supernova feedback. The actual efficiency of gas-MBHB interaction in realistic physical conditions is poorly known, and stalling of the binary is still a possibility \citep[see, e.g.,][]{Lodato2009}. In dry galaxy mergers (i.e., where gas is absent/negligible and therefore can not substantially affect the dynamics), a MBHB can evolve only because of stellar interactions. Indeed,
after the MBHs form a bound pair at $a_i\approx GM/\sigma^2$ ($M=m_1+m_2$ is the total mass of the binary, and $\sigma$ is the stellar velocity dispersion), the binary hardens by ejecting stars via single three-body interactions. The efficiency of this process saturates at the hardening radius $a_h\sim Gm_2/4\sigma^2$ \citep[where $m_2$ is the mass of the secondary,][]{Quinlan1996} and beyond that point the MBHB hardens at a constant rate. However, since in the process stars are ejected by the slingshot mechanism, the orbital decay soon falters, unless new stars are forced to replenish the otherwise depleting loss-cone. Because typically $a_h\sim 1$ pc for $\sim 10^8 \msun$ black holes, it is not guaranteed that the MBHB can eventually close the gap down to separations $a_{\rm gr}\sim 10^{-2}$ pc, i.e., the separations at which GW emission alone can drive the two MBHs to coalesce within a Hubble time. In the literature, this is often referred to as the ``final parsec problem'' \citep{Milosavljevic2003}.

In dry galaxy mergers a possible important mechanism that could solve this problem is provided by triple 
MBH interactions. Triple systems can form when a MBHB stalled at separations $\lesssim a_{\rm h}$ (because of the lack of sufficient gas and inefficient loss-cone replenishment) interacts with a third MBH -- the ``intruder'' -- carried by a new galaxy merger \citep[see, e.g.,][]{Mikkola1990,Heinamaki2001,Blaes2002,Hoffman2007,Kulkarni2012}. More specifically, these hierarchical triplets  -- i.e., triple systems where the hierarchy of orbital separations 
defines an inner and an outer binary, the latter consisting of the intruder and the centre of mass of the former -- may undergo Kozai-Lidov (K-L) oscillations \citep{Kozai1962,Lidov1962}. 
These resonances arise in the framework of a secular analysis of hierarchical triplets by Taylor-expanding the Hamiltonian of the system in powers of the inner to outer semi-major axis ratio, which is assumed to be small. The results of Kozai and Lidov, valid at the quadrupole order of approximation, demonstrate that if the intruder is on a highly inclined orbit with respect to the inner binary, the K-L mechanism tends to secularly increase the eccentricity of the inner binary, eventually driving it to coalescence. By considering a higher-order approximation (octupole), richer physics arises, and in particular significant eccentricity can build up even for triplets with low relative inclinations \citep[see e.g.,][for a comprehensive review and references]{Naoz2016}.

We have recently started a comprehensive study of post-Newtonian (PN) dynamics of MBH triplets in a cosmological framework, ultimately aiming at the full characterisation of the GW signal from the cosmic population of MBHBs. In a first step of the project \citep[][hereafter Paper I]{Bonetti2016}, 
we have discussed the importance of both chaotic three-body encounters and the K-L mechanism in the dynamics of triplets, and we have presented and tested our three-body PN code, which includes a realistic galactic potential, orbital hardening of the outer binary, the effect of the dynamical friction on the early stages of the intruder dynamics, as well
as the PN contributions to the dynamics (through 2PN order in the conservative sector, and leading order in the dissipative one). As detailed in Paper I, the employed equations of motion are consistently derived from the three-body PN Hamiltonian, which, for the first time in the framework of MBH interactions, allows us to take into account the effect of the PN three-body terms in the dynamics. Indeed, in the common practice, these kind of terms are usually not accounted for, as the two-body PN corrections are simply applied to each pair of bodies \citep[see, e.g.,][]{Mikkola2008,Rantala2017,Ryu2017}.
Finally, in a spin-off effort \citep{Bonetti2017a}, we have highlighted and solved some subtle problems affecting naive implementations of quadrupolar and octupolar gravitational waveforms from numerically-integrated trajectories of three-body systems.

In the present paper, formally the second of the series, we will perform a systematic study of the dynamics and evolution of MBH triplets. Employing the code presented in Paper I, we will explore a large region of the 6-dimensional parameter space of these systems, in terms of MBH masses, eccentricities and relative inclinations. This will allow us to fully characterise the evolution of MBH triplets in galactic potentials. In a companion paper \citep{Bonetti2017c}, we will then frame the whole picture in the hierarchical build-up of cosmic structures \citep{Barausse2012}, adopting a semi-analytic model of galaxy and MBH co-evolution, and  asses the contribution of MBH triplets to the stochastic background of nHz GWs.
In a follow-up paper (Bonetti et al. in preparation), we will also explore the implications for the GW signal from MBHBs in the mHz regime targeted by LISA.

The paper is organised as follows: in Section~\ref{sec:ICs} we describe the computational setup used to perform the simulations; in Section~\ref{sec:results} we present the results of our analysis, while in Section~\ref{sec:discusssion} we discuss in more detail the strengths and caveats of our work. Finally, in the last Section, we draw our conclusions. Throughout the paper $G$ and $c$ represent Newton's gravitational constant and the speed of light, respectively.

%%%%%%%%%%%%%%%%%%%%%%%%%%%%%%%%%%%%%%%%%%%%%%%%%%%%%%%%%%%%%%%%%%%%%%%%%%%%%%%%%%%%
%%%%%%%%%%%%%%%%%%%%%%%%%%%%%%%%%%%%%%%%%%%%%%%%%%%%%%%%%%%%%%%%%%%%%%%%%%%%%%%%%%%%
\section{Methodology}
\label{sec:ICs}

We numerically integrate the orbits of MBH triplets formed by a stalled MBHB at the centre of a stellar spherical potential ($m_1,\,m_2$), and by a third MBH ($m_3$) approaching the system from larger distances. 

\subsection{Galactic potential and scaling relations}

The properties of the stellar distribution are detailed in Paper I (to which we refer for more details). Here, we only provide a brief summary of the adopted setup.

The host's stellar distribution is modelled as a Hernquist profile with total mass $M_\star$ and scale radius $r_0$ \citep{Hernquist1990}. Moreover, in order to mimic the erosion of the nuclear region caused by the stalled inner binary \citep[see, e.g.,][]{Ebisuzaki91,VolonteriCores2003,Merritt2006,Antonini2015,Antonini_Barausse2015}, we assume the presence of a shallow central core with scale length $r_c$. With the above assumptions the chosen density profile reads 

%%%%%%%%%%%%%%%%%%%
\begin{equation}\label{eq:density}
	\rho(r) = 
	\begin{cases} 
	\dfrac{M_\star}{2\pi} \dfrac{r_0}{r_c(r_c+r_0)^3} \left(\dfrac{r}{r_c}\right)^{-1/2}       & \text{if } r \leq r_c,
	\ \\
	\dfrac{M_\star}{2\pi} \dfrac{r_0}{r(r+r_0)^3}& \text{if } r > r_c.
	\end{cases}
\end{equation}
%%%%%%%%%%%%%%%%%%%

The numerical values of the total mass, the central velocity dispersion and the scale radius are then consistently determined from empirical scaling relations that link these quantities to the hosted MBH's mass \citep[see, e.g.,][]{Dabringhausen2008,Kormendy2013}. In particular, once the total MBH mass has been fixed ($M = m_1+m_2$), following \citet{Sesana2015} the stellar mass $M_{\star}$ and the velocity dispersion $\sigma$ can be obtained from 
%%%%%%%%%%%%%%%%%%%%%%%%%%
\begin{equation}\label{eq:M_bulge}
\dfrac{M}{10^9 \rm M_{\odot}} = 0.49 \left(\dfrac{M_{\star}}{10^{11}\rm M_{\odot}}\right)^{1.16},
\end{equation}
%%%%%%%%%%%%%%%%%%%%%%%%%%
and
%%%%%%%%%%%%%%%%%%%%%%%%%%
\begin{equation}\label{eq:M_sigma}
\dfrac{M}{10^9 \rm M_{\odot}} = 0.309 \left(\dfrac{\sigma}{200 \ \rm km/s}\right)^{4.38}. 
\end{equation}  
%%%%%%%%%%%%%%%%%%%%%%%%%	
The scale radius $r_0$ can be derived from the galaxy effective radius $R_{\rm eff}$, which, unlike $r_0$, can be constrained from observations. \citet{Dabringhausen2008} found that for elliptical galaxies $R_{\rm eff}$ can be described as 
%%%%%%%%%%%%%%%%%%%%%%%%%%
\begin{equation}\label{eq:R_eff_obs}
\dfrac{R_{\rm eff}}{\rm pc} = {\rm max} \left[2.95 \left(\dfrac{M_{\star}}{10^6 \msun}\right)^{0.596}, \ 34.8 \left(\dfrac{M_{\star}}{10^6 \msun}\right)^{0.399}\right],
\end{equation}
%%%%%%%%%%%%%%%%%%%%%%%%%%
which, combined with $R_{\rm eff} \approx 1.81 r_0$ \citep{Dehnen1993}, gives the scale radius of the Hernquist profile.

Finally, the shallow density core is determined by accounting for the mass deficit caused by the binary, quantified following \citet{Merrit2013book} and \citet{Antonini2015,Antonini_Barausse2015}:
%%%%%%%%%%%%%%%%%%%%
\begin{equation}\label{eq:mass_def}
\Delta M = M \left[0.7 q^{0.2} + 0.5\ln\left(0.178 \dfrac{c}{\sigma}\dfrac{q^{4/5}}{(1+q)^{3/5}}\right)\right],
\end{equation}
%%%%%%%%%%%%%%%%%%%%
where $q=m_2/m_1$. The core radius $r_c$ is then obtained by imposing that $\Delta M$ equals the mass difference, within $r_c$, between the original Hernquist profile and the $r^{-1/2}$ profile (cf. eq.~\ref{eq:density}).

We do not consider any dark matter (DM) extended component associated to the stellar background. The properties of the stellar background are kept fixed during the evolution of the MBH triplet, so that its properties are solely determined by the mass of the stalled MBHB through the scaling relations discussed above.

%%%%%%%%%%%%%%%%%%%%%%%%%%%%%%
\begin{figure}
	\centering
	\includegraphics[scale=0.28]{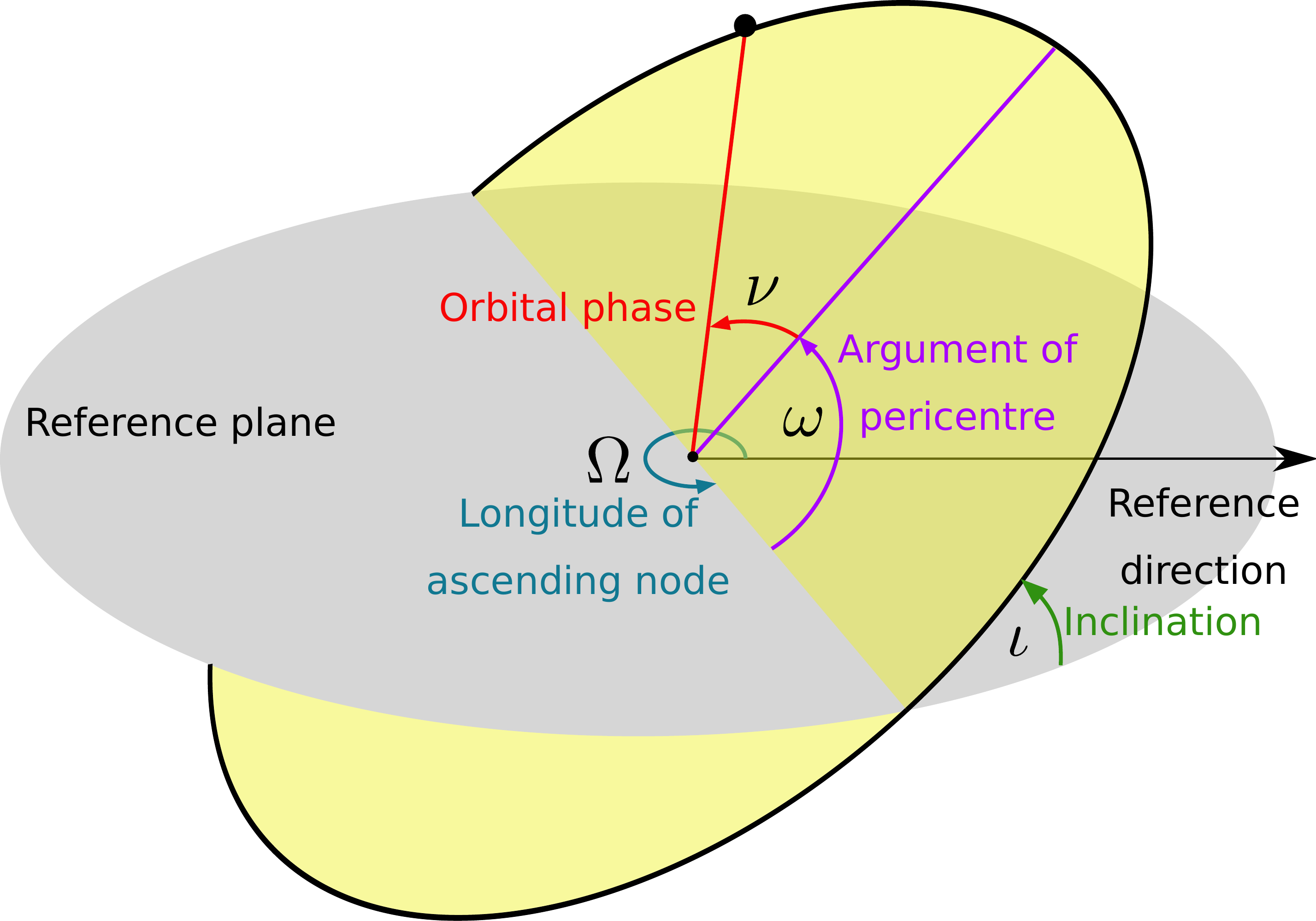}
	\caption{Schematic representation of a binary orbit in 3D space.}
	\label{fig:binary3d}
\end{figure}
%%%%%%%%%%%%%%%%%%%%%%%%%%%%%%

\subsection{Initial orbital parameters}

In principle, the complete characterisation of a system of three MBHs requires specifying 21 parameters. However, the initial configuration of the considered systems allows us to reduce the number of free parameters.

We first fix the motion of the centre of mass (position and velocity), thus reducing the effective number of initial parameters to 15. We then note that the initial values of some orbital elements (see figure~\ref{fig:binary3d}) do not play any major role in the dynamics. More specifically, among these we have the two arguments of pericentre, the two longitudes of the ascending node, and the two orbital phases. Indeed, the various physical processes (e.g., dynamical friction, stellar hardening, the precession due to the galactic potential and the relativistic one), driving the evolution of MBHBs, quickly make these parameters ``lose memory'' of the their initial values, as soon as the three MBHs first bind in a hierarchical triplet. The number of considered initial parameters is then reduced to 9. Moreover, we set the initial semi-major axis of the $m_3$ orbit to the scale radius $r_0$ of the stellar bulge, which ultimately depends only on  $m_1+m_2$, thus reducing the number of relevant free parameters to 8. We have verified that our results are robust against this choice of initial semi-major axis.

We note that the choice to initialise the inner binary as stalled implicitly determines its separation. In fact, according to a vast literature on the evolution of MBHBs \citep[see, e.g.,][]{Begelman1980,Saslaw1974,Quinlan1996,Yu2002,Sesana2006} and the final parsec problem \citep{Milosavljevic2003,Vasiliev2014,Vasiliev2015}, the binary shrinks to a separation $a_{\rm in} \lesssim a_h$, where $a_h \sim G m_2/4 \sigma^2$ represents the  hardening radius of the stellar distribution. Further hardening of the system proceeds then at a nearly constant rate, dictated by the efficiency at which stars can be supplied to the binary loss cone. Depending on the properties of the host galaxy, evolution timescales can be as long as several Gyr \citep[see, e.g.,][]{Yu2002,Berczik2006,Preto2011,Khan2011,Gualandris2012b,Vasiliev2014,lastPc2,Vasiliev2015,Sesana2015,Khan2016,Gualandris2017}. Therefore, we initialise the inner binary at a separation around $a_h$, whose value, once $m_1$ and $m_2$ are specified, is completely determined. The exact value of $a_{\rm in}$ is practically irrelevant as long as it is close to $a_h$ and sufficiently larger than $a_{\rm gw}$, where $a_{\rm gw}$ represents the scale at which GW emission starts dominating the evolution of the binary. Specifically, for the initialisation of $a_{\rm in} \lesssim a_h$, we assume (somewhat arbitrarily) $a_{\rm in}/a_{\rm gw}=(a_h/a_{\rm gw})^{3/4}$, and we have checked that the results are robust against this choice. At this point, we are  left with only 7 free initial conditions.

Finally, the isotropy of the problem allows us to specify the relative inclination between the orbital planes of the two orbits, i.e., $\iota \equiv \iota_{\rm in}+\iota_{\rm out}$, thus reducing the final set of relevant initial free parameters to 6. We will explore this parameter space in the following.

%%%%%%%%%%%%%%%%%%%%%%%%%%%%%%%%%%%%%%%%%%%%%%%%
\begin{table}
	\centering
	\caption{Parameter space sampling}
	\label{tab:param_space}
	\begin{tabular}{ll}
		\hline
		\multicolumn{2}{c}{Initial conditions}\\
		\hline
		& \\
		$\log(m_1)$ [$\rm M_{\odot}$] & 5,\,6,\,7,\,8,\,9,\,10\\
		$\log(q_{\rm in})$ &  -1.5,\, -1.0,\, -0.5,\, 0.0  \\
		$\log(q_{\rm out})$ & -1.5,\, -1.0,\, -0.5,\, 0.0  \\
		$e_{\rm in}$ & 0.2,\, 0.4,\, 0.6,\, 0.8 \\
		$e_{\rm out}$ & 0.3,\, 0.6,\, 0.9 \\
		cos$\,\iota$ & 13 values equally spaced in $(-1,1)$\\
		\hline
	\end{tabular}
\end{table}
%%%%%%%%%%%%%%%%%%%%%%%%%%%%%%%%%%%%%%%%%%%%%%%%%%

In generating the initial conditions, for the mass of the heavier MBH of the inner binary ($m_1$) we choose 6 values uniformly selected in logarithmic space, from 
$10^5$ M$_\odot$ to $10^{10}$ M$_\odot$. The inner and outer binary mass ratios $q_{\rm in}\equiv m_2/m_1$ and $q_{\rm out} \equiv m_3/(m_1+m_2)$ can take 4 values each, uniformly spaced (logarithmically) from $0.03$ to $1$. The eccentricity of the inner binary, $e_{\rm in}$, takes 4 values uniformly spaced from $0.2$ to $0.8$, while the eccentricity of the outer binary, $e_{\rm out}$, is chosen among $0.3,0.6,0.9$. 

Finally, in order to average our results over an isotropic orientation of the angular momenta of the two binaries, we sample the relative inclination of the two orbital planes, $0^\circ < \iota < 180^\circ$, in 13 values equally spaced in $\cos\iota$. 

When presenting results marginalised over $e_{\rm in}$ and $e_{\rm out}$, those are simply obtained by summing up simulations with different eccentricities, which corresponds to a uniform weight in $e_{\rm in}$ and $e_{\rm out}$. Similarly, results marginalised over $q_{\rm in}$ and $q_{\rm out}$ are also obtained by direct summation, which corresponds to a uniform weight in the logarithm of the mass ratios. The sampling of the 6-dimensional space is summarised in table~\ref{tab:param_space}, and consists of a grand total of 
14,976 different initial conditions.

Simulations are run with the code presented in Paper I, which we briefly summarise here. 
The employed numerical scheme directly integrates the three-body (Hamiltonian) equations of motion through 2.5PN order (i.e. through 2PN order in the conservative dynamics and leading order in the dissipative one), introducing velocity-dependant forces to account for the dynamical friction on the intruder during its initial orbital decay toward the galactic centre, 
and for the stellar hardening \citep{Quinlan1996} of the outer binary. 
Unlike in Paper I, the centre of mass of the triplet is not re-centred every 1,000 integration steps, but we rather apply the following algorithm: when the MBH dynamics is dominated by the stellar background, dynamical friction acts on the binary {\it and} on the perturber separately. When $m_3$ later binds to the inner binary (thus forming the outer binary), the dynamical friction force is instead applied to the centre of mass of the triplet, and the stellar hardening of the outer binary is simultaneously activated. Stellar hardening is eventually switched off as the first close three-body MBH encounter occurs and the dynamics becomes chaotic. Moreover, in order speed up our computations, 
we switch off the conservative 2PN terms in the Hamiltonian dynamics. We have checked in Paper I that 2PN corrections are indeed negligible, at least in a statistical sense, although extremely time-consuming computationally.

We stop the orbital integration when one of the following conditions is first met: a minimum approach between two members of the triplet is reached;  one of the MBHs is ejected; or the time spent exceeds the (present) Hubble time. Regarding the first condition, the minimum separation is set to 15 gravitational radii, i.e., the spatial threshold is given by $15 \ G(m_i+m_j)/c^2$, where $m_i$ and $m_j$ represent the masses of the merging MBHs. When that separation is reached, we count the event as a ``binary coalescence". \footnote{The choice about a minimum approach threshold is not an option, but a real requirement in the PN framework. Indeed, given its perturbative character, it is clear that as soon as two MBHs get sufficiently close,	a full GR solution is needed to actually describe the merger event. Within the PN framework, we can safely describe the dynamics until the very last phases before the coalescence. Moreover, in order to to avoid possible unphysical behaviours determined by a not sufficiently high PN dynamics (we have switched off the 2PN terms because of their huge computational cost), we have chosen such a conservative threshold.} An ejection, instead, is counted whenever one of the MBHs is kicked to a distance in excess of 10 stellar bulge scale radii, irrespective of its binding energy. 
Note that this threshold is rather conservative compared to, e.g., \citet{Hoffman2007}, and has been chosen to avoid overestimating the interaction rate between the inner binary and the returning kicked MBH. Indeed, in a perfect spherically symmetric potential like ours, an MBH bound to the galaxy potential would always return to the centre of the stellar distribution. In more realistic situations, however, any deviation from spherical symmetry would prevent further interactions of the kicked MBH with the inner binary \citep[see, e.g.,][]{Guedes2009}. Our combined choices of, i) neglecting the DM component of the galactic potential, and ii) counting kicked MBHs as ejected once they reach a relatively short distance from the centre, are then conservative in terms of predicted MBHB coalescences. We plan to analyse in details the effects of triaxiality on the dynamics of MBH triple systems in the future.

%%%%%%%%%%%%%%%%%%%%%%%%%%%%     
\begin{table}
	\centering
	\caption{Merger percentage}
	\label{tab:merger_summary}
	\begin{tabular}{ccccl}
		\hline
		$\log m_1$ & \multicolumn{4}{c}{\% Mergers} \\
		$\rm [M_{\odot}]$ & $m_1$-$m_2$  & $m_1$-$m_3$  & $m_2$-$m_3$ & Total\\
		\hline
		& \\
		5  & 16.8 & 0.9 & 0.8 & 18.5(1.6)  \\
		6  & 16.2 & 1.4 & 1.0 & 18.5(1.9)  \\
		7  & 15.4 & 2.5 & 1.4 & 19.4(4.4)  \\
		8  & 14.7 & 4.0 & 2.5 & 21.2(6.3)  \\
		9  & 15.2 & 4.1 & 3.2 & 22.5(11.2) \\
		10 & 21.1 & 7.6 & 3.3 & 31.9(12.7) \\
		\hline
	\end{tabular}
\end{table}      
%%%%%%%%%%%%%%%%%%%%%%%%%%%%%%  

%%%%%%%%%%%%%%%%%%%%%%%%%%%%%%%%%%%%%%%%%%%%%%%%%%%%%%%%%%%%%%%%%%%%%%%%%%%%%%%%%%%%
%%%%%%%%%%%%%%%%%%%%%%%%%%%%%%%%%%%%%%%%%%%%%%%%%%%%%%%%%%%%%%%%%%%%%%%%%%%%%%%%%%%%
\section{Results}
\label{sec:results}

\subsection{Merger fraction}

Our full results, in terms of merger fractions as functions of different triplet parameters, are reported 
in a series of tables presented in Appendix \ref{sec:app_A}.

Table~\ref{tab:merger_summary} shows in particular the dependence of the merger fraction, i.e. the fraction of simulations ending with a merger of any two members of the triplet, on the  mass of the primary MBH ($m_1$). As can be seen, the merger fraction is almost constant and around $\simeq 20\%$ for the entire sampled mass range, except for the most massive case, where $\gsim 30\%$ of the systems are bound to coalescence. Averaged over $m_1$, the merger fraction is $\simeq 22\%$. The merger excess for $m_1 = 10^{10}$ M$_\odot$  is most probably due to the way we generate the initial conditions. Since the inner binary is initialised with a separation of the order of its hardening radius, $a_{h} = G m_2/(4\sigma^2)$, and since the efficiency of GW emission scales with the binary mass, high-mass/low-$q_{\rm in}$ systems are not technically stalled. Indeed, their coalescence timescale under GW emission, albeit of several Gyr, is still shorter than the  Hubble time~\citep{Sesana2010,Dvorkin2017}.

In table~\ref{tab:merger_summary} we also report, as an ancillary entry in the  column ``Total", the fraction of MBHBs that are bound 
to coalesce within a Hubble time {\it after} an ejection event.  Note that since we stop our simulations whenever an ejection occurs, we compute {\it a posteriori}
 the time the remaining MBHB needs to coalesce because of GW losses. These ``post-ejection'' coalescences add a further $\simeq 6\%$ to the overall merger fraction (hence accounting for $\simeq 1/5$ of the total number of mergers), 
which is then $\simeq 30\%$. Taken at face values, our results confirm that triple interactions represent a possible, albeit partial, solution to the final-parsec problem. 

%%%%%%%%%%%%%%%%%%%%%%%%%%%%%%
\begin{figure}
	\centering
	\includegraphics[scale=0.31]{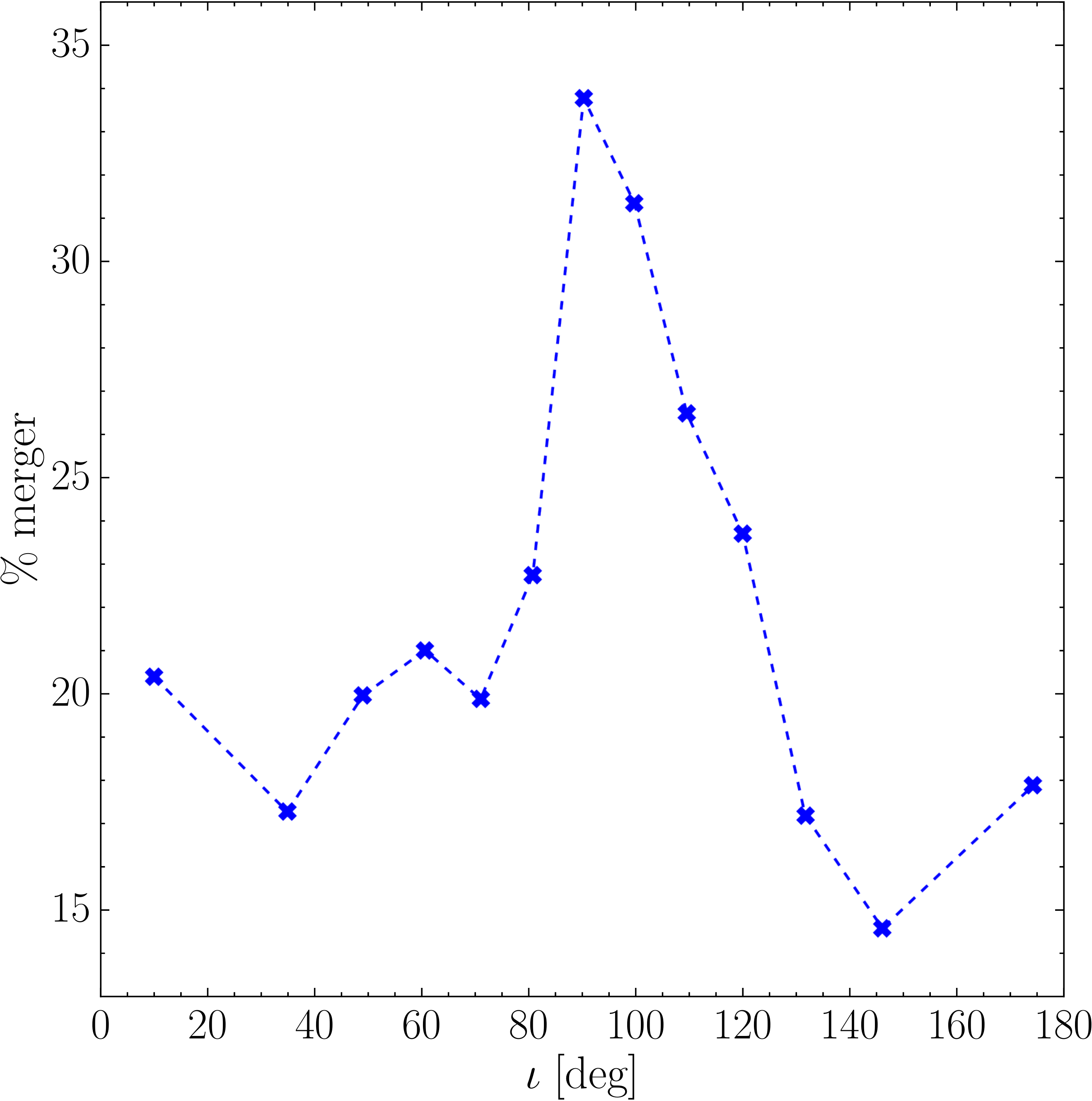}
	\caption{Merger fraction as a function of the initial relative inclination between the inner and outer binary. Post-ejection coalescences are not included. The prominent peak at $\iota \approx 90^\circ$ confirms the important role of the  K-L mechanism in driving the merger of MBHBs. }
	\label{fig:merger_fraction_all_mass_inc}
\end{figure}
%%%%%%%%%%%%%%%%%%%%%%%%%%%%%%

%%%%%%%%%%%%%%%%%%%%%%%%%%%%%%
\begin{figure}
	\centering
	\includegraphics[scale=0.33]{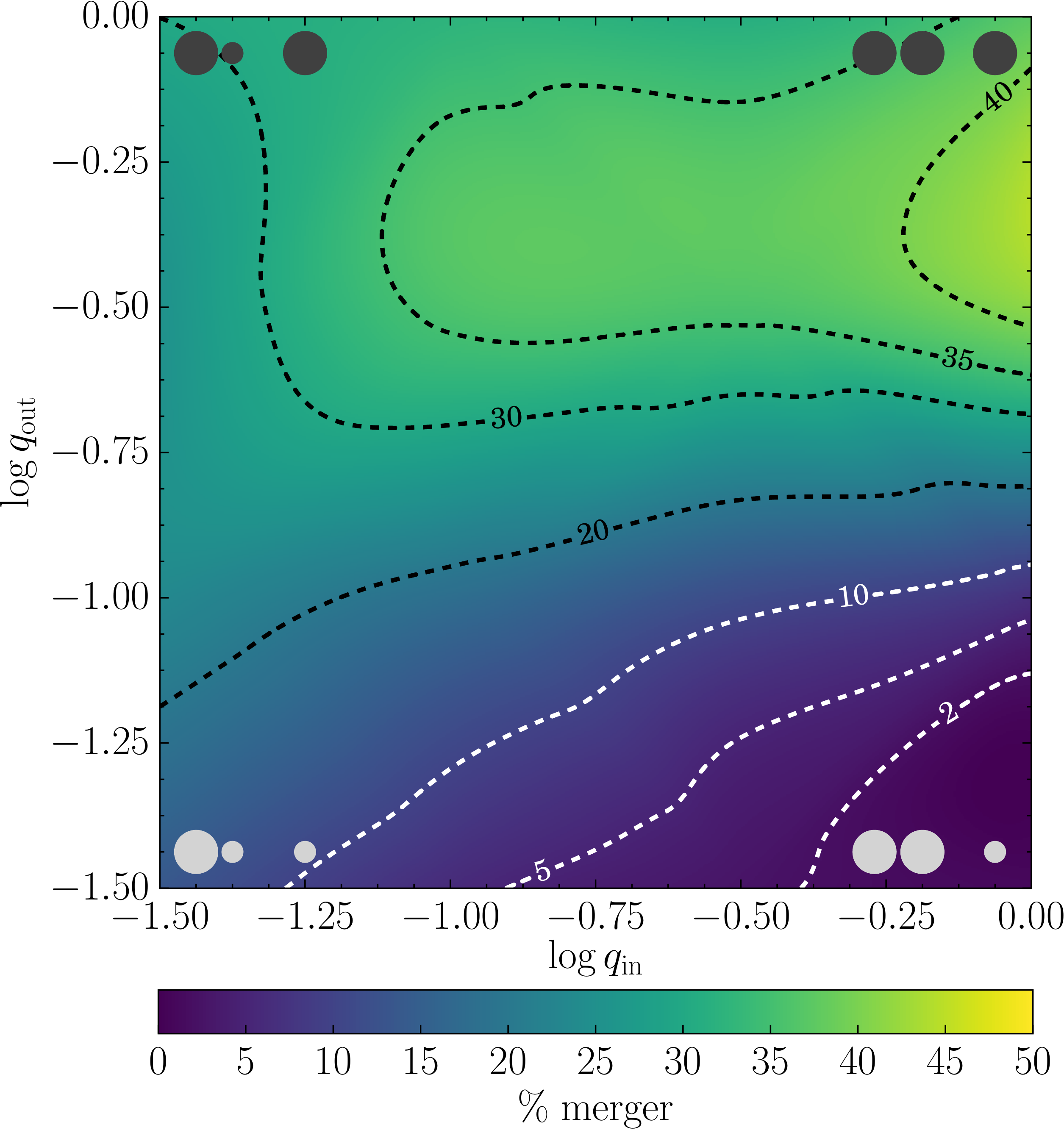}
	\caption{Merger fraction (colour coded) as a function of the inner ($q_{\rm in}=m_2/m_1$) and outer ($q_{\rm out}=m_3/(m_1+m_2)$) mass ratios. The merger fraction is larger ($\gsim 30$ \%) in the upper region, corresponding to $q_{\rm out} \gsim 0.3$ and $0.1 \lsim q_{\rm in} \lsim 1$. The circles in the four corners of the plot represent, in cartoon-like fashion, the corresponding mass hierarchy of the triplets.}
	\label{fig:merger_fraction_mass_ratio}
\end{figure}
%%%%%%%%%%%%%%%%%%%%%%%%%%%%%%

In figure~\ref{fig:merger_fraction_all_mass_inc} the merger fraction ({\it not} inclusive of the post-ejection coalescences discussed above) is plotted as a function of the initial relative inclination of the two binaries. The merger fraction peaks around $\simeq 90^\circ$, which is indeed the angle yielding the maximal eccentricity excitation 
in the standard (i.e., quadrupole-order) K-L mechanism. K-L oscillations have therefore a strong impact on the dynamics of our simulated MBHBs.

In a two-dimensional map (figure~\ref{fig:merger_fraction_mass_ratio}) we show again the merger fraction, but now as a function of the initial values of $q_{\rm in}$ and $q_{\rm out}$\footnote{Note that an equal-mass triplet (i.e., $m_1=m_2=m_3$) is characterised by $\log q_{\rm in}=0$ and $\log q_{\rm out}=-0.3$.}. The peak of the merger fraction occurs for equal-mass triplets, but there is a large plateau in the upper part of the plot, with hints of two distinct maxima. A simple interpretation is that the inner binary, in order to be perturbed, needs to interact 
with an intruder of at least comparable mass (i.e., $ \log q_{\rm out} \approx -0.3$). A light $m_3$ is most probably simply kicked out by the heavier inner binary, as hinted at by the rapid decline of the merger fraction as $q_{\rm out}$ gets  $\ll 1$ \citep[see,][]{Heggie1975}. Note, however, that even for low values of $q_{\rm out}$ 
 the merger fraction is significant when $q_{\rm in}$ is also small (thus, when $m_2\approx m_3$). 
 
%%%%%%%%%%%%%%%%%%%%%%%%%%%%%
\begin{figure}
	\centering
	\includegraphics[scale=0.305]{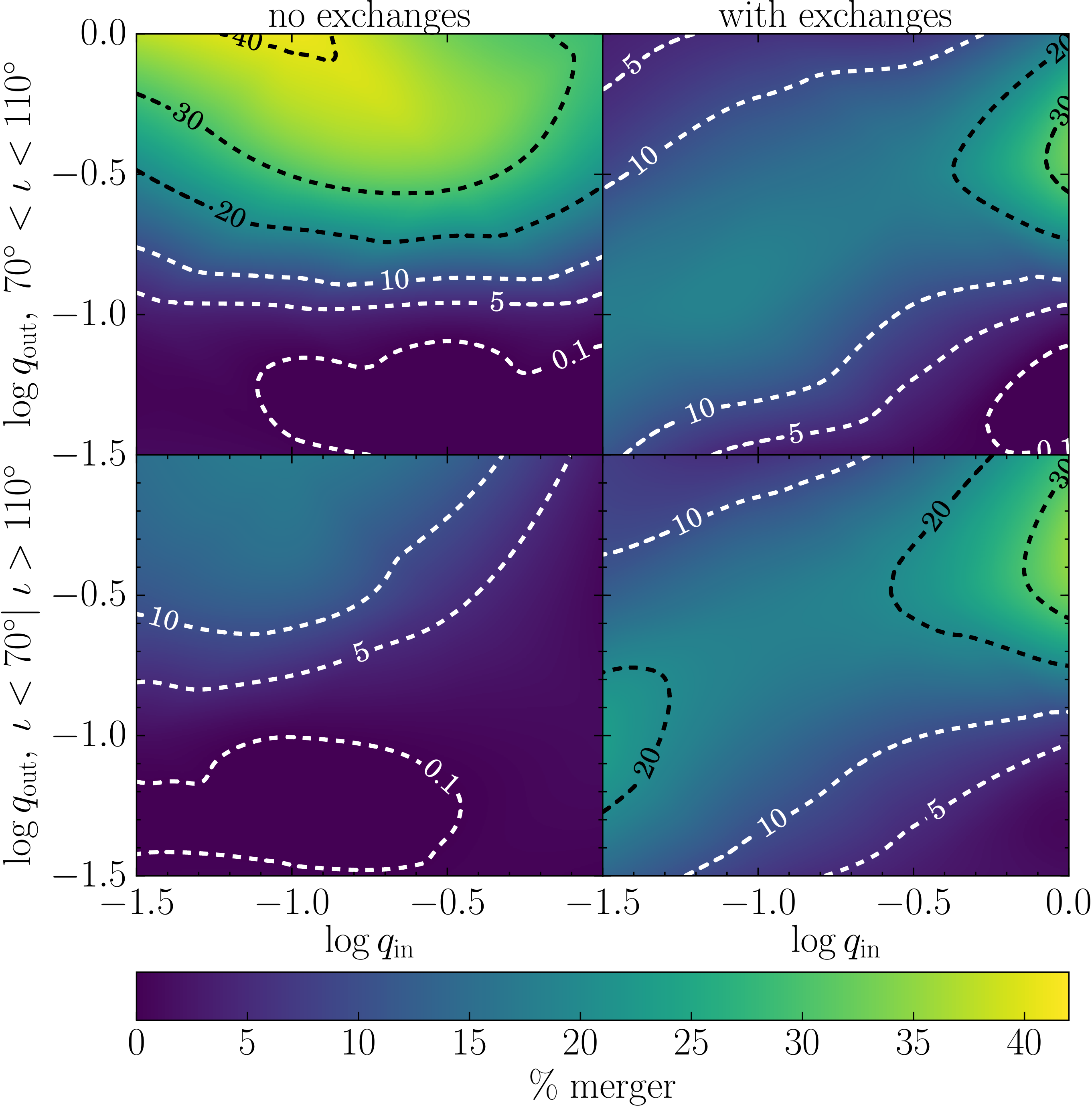}
	\caption{Break-up of figure \ref{fig:merger_fraction_mass_ratio} into different sub-populations. The two columns distinguish between merging binaries that underwent at least one close encounter leading to an exchange (right) and merging binaries that experienced no such encounters (left). The two rows identify sub-populations starting off with relative inclinations $70^\circ<\iota<110^\circ$ (top) and $\iota<70^\circ$ or $\iota>110^\circ$ (bottom). Merger fractions are normalized with respect to the total number of simulations in the respective inclination range.}
	\label{fig:merger_fraction_mass_ratio_bin}
\end{figure} 
%%%%%%%%%%%%%%%%%%%%%%%%%%
 
%%%%%%%%%%%%%%%%%%%%%%%%%%%%%
\begin{figure}
	\centering
	\includegraphics[scale=0.33]{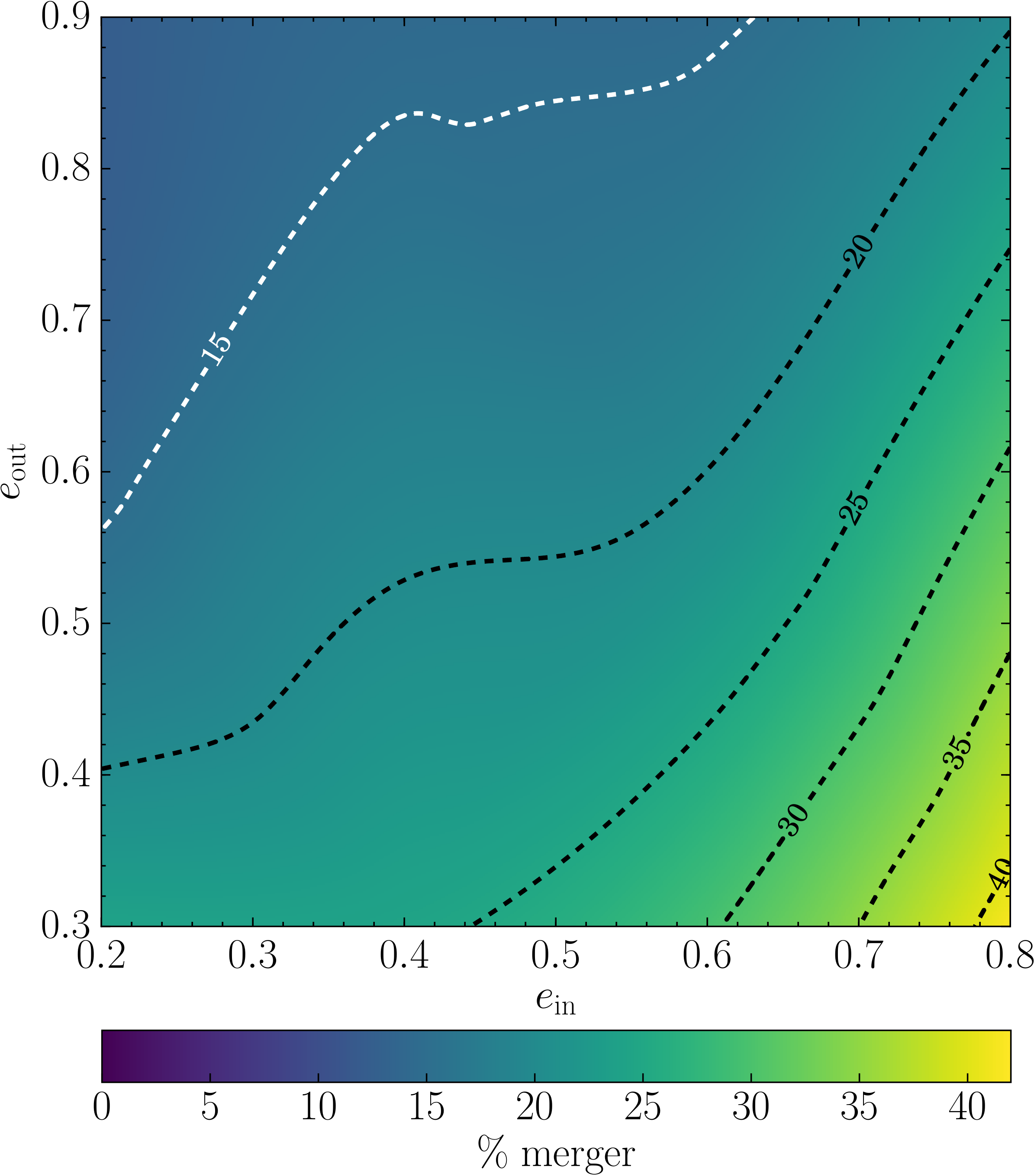}
	\caption{Merger fraction (colour coded) as a function of the initial inner ($e_{\rm in}$) and outer ($e_{\rm out}$) binary eccentricities.}
	\label{fig:merger_fraction_ein_eout}
\end{figure} 
%%%%%%%%%%%%%%%%%%%%%%%%%%
 
 A finer understanding of the results of figure~\ref{fig:merger_fraction_mass_ratio} can be gained by considering that, at the octupole level, the K-L oscillations are more easily triggered when the inner binary has a small mass ratio. In this case, the inner binary can merge when the triplet is still in the initial hierarchical phase, and the eccentricity growth responsible for the coalescence is primarily driven by secular processes. This is the cause of the leftmost peak in the merger fraction, indeed occurring for $\log q_{\rm out} \approx -0.3$ and $\log q_{\rm in} \ll 0$. A second channel to coalescence is represented by merger-inducing strong non-secular close encounters that the original inner binary 
experiences once the triplet becomes unstable. This happens for almost equal-mass triplets, i.e., when the intruder carries a mass sufficiently large to perturb the inner binary, but, at the same time, the K-L mechanism is not easily triggered.  During the process, prior to coalescence, several exchanges\footnote{An exchange is an event in which the intruder kicks one body (usually the lightest one) out of the inner binary, and binds to the other to form a new two-body system.} may occur, and therefore the final merger does not necessarily involve the members of the original inner binary. This second channel is responsible for the rightmost peak of the merger fraction in figure~\ref{fig:merger_fraction_mass_ratio}.

In order to better understand the role played by these two different channels in the merger fraction, we separately analyse the systems in which no exchange occurs during the evolution, and the rest of the systems that instead experience strong encounters, ultimately leading to one or more exchange events. In addition, we single out systems with initial inclination in the range ($70^\circ<\iota<110^\circ$), hence dividing our simulations into four subsets. 

%%%%%%%%%%%%%%%%%%%%%%%%%%%%%%%%%%%%%%
\begin{table*}
	\centering
	\caption{Comparison of the merger fraction from the simulations with $m_1 = 10^9$ M$_\odot$, run with and without conservative 1PN corrections.}
	\label{tab:cfr_PN}
	\begin{tabular}{l c c c l | c c c c c l}
		\hline
		$m_1 = 10^{9} \ \rm M_{\odot}$ & \multicolumn{4}{c|}{\% Mergers including 1PN} & & & \multicolumn{4}{c}{\% Mergers without 1PN}\\
		$q_{\rm in}/q_{\rm out}$	 & $m_1$-$m_2$  & $m_1$-$m_3$  & $m_2$-$m_3$ & Total & & & $m_1$-$m_2$  & $m_1$-$m_3$  & $m_2$-$m_3$ & Total \\
		\hline\\
		0.0316/0.0316 &   9.0 &   5.1 &   0.0 &   14.1(20.5) & & & 44.2 &   5.8 &   0.0 &   50.0(5.1 ) \\
		0.0316/   0.1 &  23.7 &   3.2 &   0.0 &   26.9(13.5) & & & 59.0 &   1.9 &   0.0 &   60.9(5.1 ) \\
		0.0316/0.3162 &   9.0 &   1.3 &   0.6 &   10.9(3.8 ) & & & 41.0 &   3.2 &   0.0 &   44.2(3.8 ) \\
		0.0316/     1 &  25.0 &   0.0 &   1.9 &   26.9(0.0 ) & & & 62.2 &   0.0 &   0.6 &   62.8(0.0 ) \\
		0.1/0.0316 &   4.5 &   1.9 &   0.0 &    6.4(16.7) & & & 29.5 &   1.9 &   0.0 &   31.4(12.8) \\
		0.1/   0.1 &  16.0 &   5.8 &   0.6 &   22.4(17.3) & & & 35.9 &   6.4 &   1.3 &   43.6(17.9) \\
		0.1/0.3162 &  24.4 &   3.2 &   1.3 &   28.8(11.5) & & & 44.9 &   4.5 &   0.6 &   50.0(7.7 ) \\
		0.1/     1 &  25.6 &   0.0 &   7.1 &   32.7(1.3 ) & & & 40.4 &   0.6 &   7.7 &   48.7(0.0 ) \\
		0.3162/0.0316 &   2.6 &   0.6 &   0.0 &    3.2(7.1 ) & & &  9.0 &   0.0 &   0.6 &    9.6(5.8 ) \\
		0.3162/   0.1 &  12.8 &   2.6 &   0.6 &   16.0(14.1) & & & 17.3 &   5.8 &   0.0 &   23.1(9.6 ) \\
		0.3162/0.3162 &  26.3 &  14.1 &   3.8 &   44.2(16.0) & & & 30.1 &  13.5 &   4.5 &   48.1(16.0) \\
		0.3162/     1 &  17.3 &   1.9 &   4.5 &   23.7(5.8 ) & & & 29.5 &   3.2 &   5.8 &   38.5(6.4 ) \\
		1/0.0316 &   1.3 &   0.0 &   0.0 &    1.3(5.1 ) & & &  3.2 &   0.6 &   0.0 &    3.8(1.3 ) \\
		1/   0.1 &   5.8 &   0.0 &   0.6 &    6.4(8.3 ) & & & 18.6 &   1.3 &   1.3 &   21.2(10.3) \\
		1/0.3162 &  26.9 &  12.2 &  14.1 &   53.2(16.0) & & & 33.3 &   9.6 &   6.4 &   49.4(13.5) \\
		1/     1 &  13.5 &  14.1 &  15.4 &   42.9(21.8) & & & 19.9 &  16.0 &  18.6 &   54.5(11.5) \\
		\hline
		Average		  &  15.2 &   4.1 &   3.2 &   22.5(11.2) & & & 32.4 &   4.6 &   3.0 &   40.0(7.9 ) \\
		\hline
	\end{tabular}
\end{table*}
%%%%%%%%%%%%%%%%%%%%%%%%%%%%%%%%%%%%

The relative merger fraction of these subsets (i.e., relative to the number of simulations performed in a particular inclination range) is shown in figure~\ref{fig:merger_fraction_mass_ratio_bin}. Left panels represent the systems in which no exchange occurs. In these cases the coalescence is mainly due to secular K-L oscillations, a fact confirmed by the comparatively much higher merger fraction (factor of $\approx3$) at high initial inclinations (upper left panel). Moreover, we note that low $q_{\rm in}$ are more likely to lead to mergers, irrespective of the inclination. As already mentioned, this is to be ascribed to the octupole terms of the K-L resonances, whose amplitude is proportional to the mass difference $m_1-m_2$, hence vanishing for $q_{\rm in}\rightarrow 1$. Note that, unlike in the standard quadrupole K-L resonances, the introduction of the octupole terms can excite high eccentricities even at low inclinations, a fact responsible for the non-negligible merger fraction in the lower left panel of figure~\ref{fig:merger_fraction_mass_ratio_bin}. At high inclinations, the region of both low $q_{\rm out}$ and low $q_{\rm in}$ should be prone to the K-L mechanism, but
 our results show no significant merger fraction (figure~\ref{fig:merger_fraction_mass_ratio_bin}, upper left panel). This can be understood by noting that if the timescale of the relativistic precession is shorter than that of the K-L oscillations, the latter are damped. Since the K-L timescale increases as $m_3$ decreases, in the limit $q_{\rm out} \ll 1$ the process is largely suppressed by relativistic precession. The relatively large merger fraction visible in the lower left area of figure~\ref{fig:merger_fraction_mass_ratio} is then due to non-secular processes.

%%%%%%%%%%%%%%%%%%%%%%%%%%%%%%%%%%%%%
\begin{figure*}
	\centering
	\includegraphics[scale=0.35]{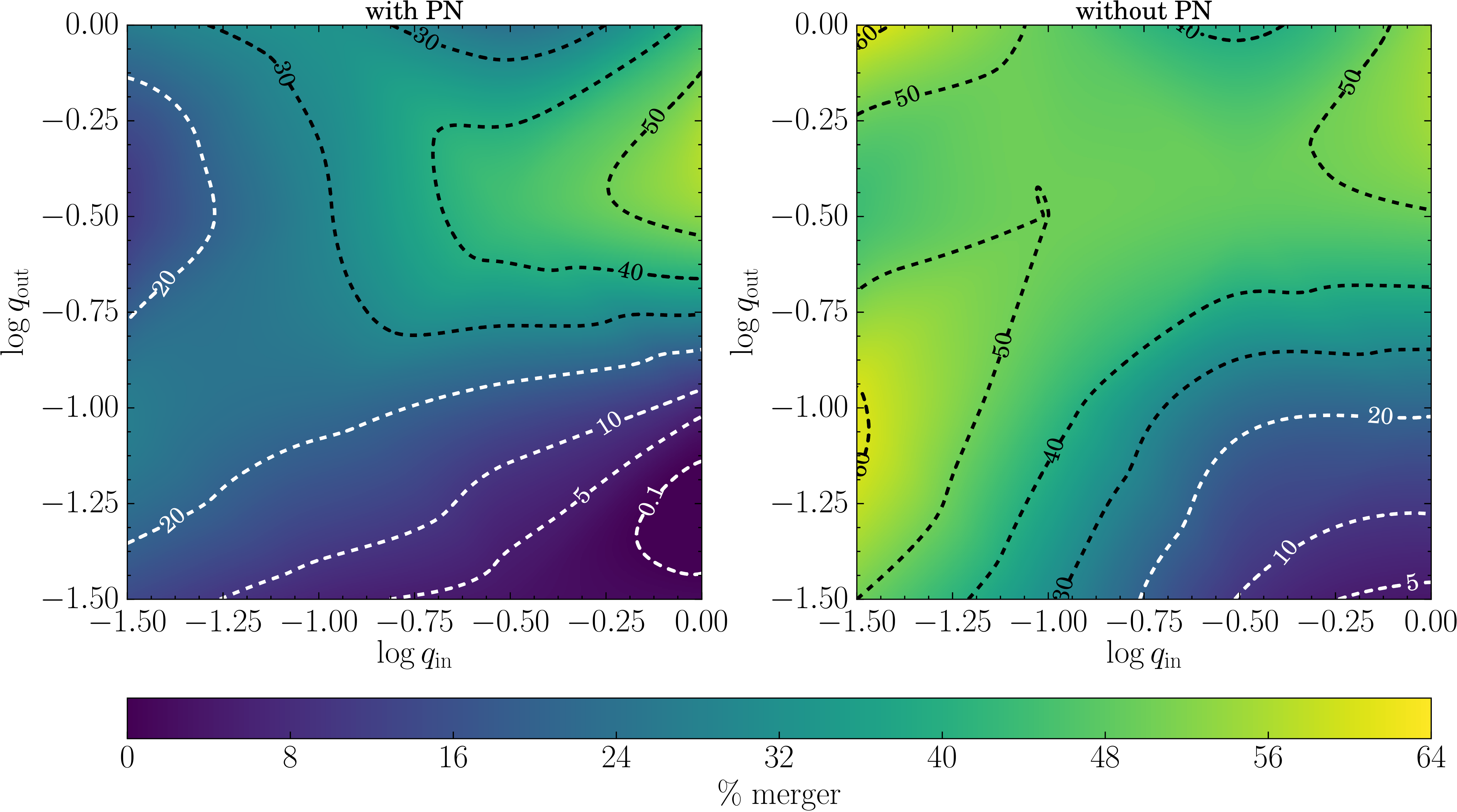}
	\caption{Comparison between the merger fraction in the case $m_1 = 10^9$ M$_\odot$, when the dynamics is evolved with (left panel) and without (right panel) 1PN term. The suppression of the merger fraction for low $q_{\rm in}$, due to 1PN precession, is clearly visible (see discussion in main text).}
	\label{fig:cfr_w_wo_PN}
\end{figure*}
%%%%%%%%%%%%%%%%%%%%%%%%%%%%%%%%%%%%%

The right panels of figure~\ref{fig:merger_fraction_mass_ratio_bin} show the non-secular channel to merger. The first thing to notice is that the pattern of the merger fraction is almost independent of the inclination angle, consistent with the fact that exchanges occur when chaotic interactions take place and secular processes play no significant role. The merger fraction is larger when the three MBHs have similar masses, and in general has non-negligible values ($> 10\%$) only along a broad band stretching from the upper right to the lower left sides of the $q_{\rm in}-q_{\rm out}$ plane. This can be understood by considering that when $q_{\rm in}\simeq 1$ and $q_{\rm out}\ll 1$ (i.e., in the lower right corner of the plot), the intruder cannot perturb significantly the much more massive inner binary. On the other extreme (i.e.,  $q_{\rm in}\ll 1$ and $q_{\rm out}\simeq 1$, upper left  corner of the plot), $m_3$ simply kicks the much lighter $m_2$ out of the inner binary, taking its place. It is only when $m_3 \sim m_2$ that genuinely chaotic dynamics can take place, in some cases leading to coalescence. 

Finally, figure \ref{fig:merger_fraction_ein_eout} shows the merger fraction as a function of the initial eccentricity of the inner and outer binary. We note that the merger fraction increases with increasing $e_{\rm in}$, while it decreases with increasing $e_{\rm out}$. The dependence upon $e_{\rm in}$ is readily understood, since highly eccentric inner binaries are closer to the efficient GW-emission stage and can easily be driven to coalescence by a relatively mild perturbation from a third body. The dependence upon $e_{\rm out}$ is likely due to the fact that quasi-circular outer binaries form a stable hierarchical triplet for a comparatively longer time during the inspiral of $m_3$, hence leaving more room to the development of K-L resonances, which are efficient at driving the inner binary to coalescence. Conversely, in very eccentric outer binaries, $m_3$ soon interacts with the inner binary at pericentre, entering the chaotic phase. Chaotic interactions are more likely to result in ejections rather than mergers, hence suppressing the overall merger fraction.

%%%%%%%%%%%%%%%%%%%%%%%%%%%%%%%%%%%%%%%%%%%%%%%%%%%%%%%%%%%%%%%%%%%%%%%%%%%%%%%%%%%%
\subsubsection{Importance of PN corrections}

As pointed out in the previous section, the K-L mechanism can be suppressed by general relativistic effects. In particular,  
relativistic precession tends to destroy the coherent pile-up of the perturbation that the third body induces on the inner binary, hence effectively damping the K-L resonances\footnote{More precisely, any kind of precession tends to suppress the K-L mechanism.}. In order to quantify the impact of the relativistic precession on the merger fraction and to compare our results with previous work that neglected this effect \citep[e.g.][]{Iwasawa2006,Hoffman2007}, in table~\ref{tab:cfr_PN} and in figure~\ref{fig:cfr_w_wo_PN} we compare, only for the case $m_1=10^9$ M$_\odot$, the merger fraction obtained with and without 1PN corrections. Overall, the merger fraction is substantially higher in the case without 1PN terms (right panel of figure~\ref{fig:cfr_w_wo_PN}). As can be seen, the largest differences compared to the full case occur for $q_{\rm in}\ll 1$, where, because of octupole-order terms, the K-L mechanism is maximally effective. On the contrary, for large $q_{\rm in}$ the merger fractions with or without 1PN corrections are similar, because, as previously discussed, coalescences are mainly due to chaotic strong encounters, rather than to K-L oscillations. Our results highlight the importance of K-L resonances in inducing MBH mergers in triple systems, and the need to account at least for 1PN corrections. 

%%%%%%%%%%%%%%%%%%%%%%%%%%%%%%%%
\subsection{Merger timescales}

%%%%%%%%%%%%%%%%%%%%%%%%%%%%%%
\begin{figure}
	\centering
	\includegraphics[scale=0.31]{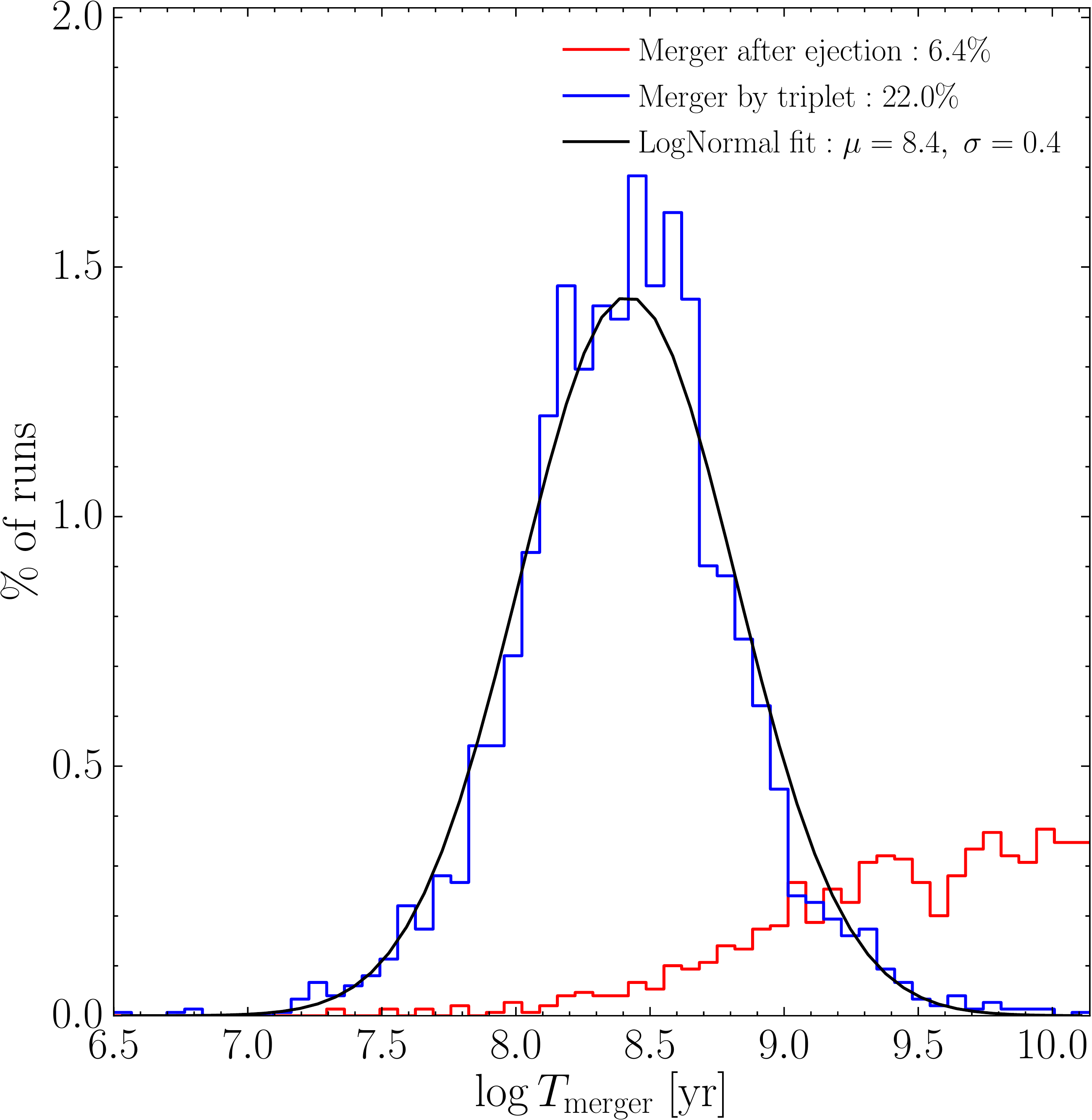}
	\caption{Distribution of merger-times. The blue histogram represents binaries that merge because of the prompt interaction with a third body (chaotic dynamics and/or K-L resonances plus GWs), and the black line is a log-normal fit to the distribution. The red histogram, instead, represents the binaries that are driven to merger by GW emission alone after the ejection of one of the MBHs (always the lightest one).}
	\label{fig:T_merger_all}
\end{figure}
%%%%%%%%%%%%%%%%%%%%%%%%%%%%%%

%%%%%%%%%%%%%%%%%%%%%%%%%%%%%%
\begin{figure}
	\centering
	\includegraphics[scale=0.31]{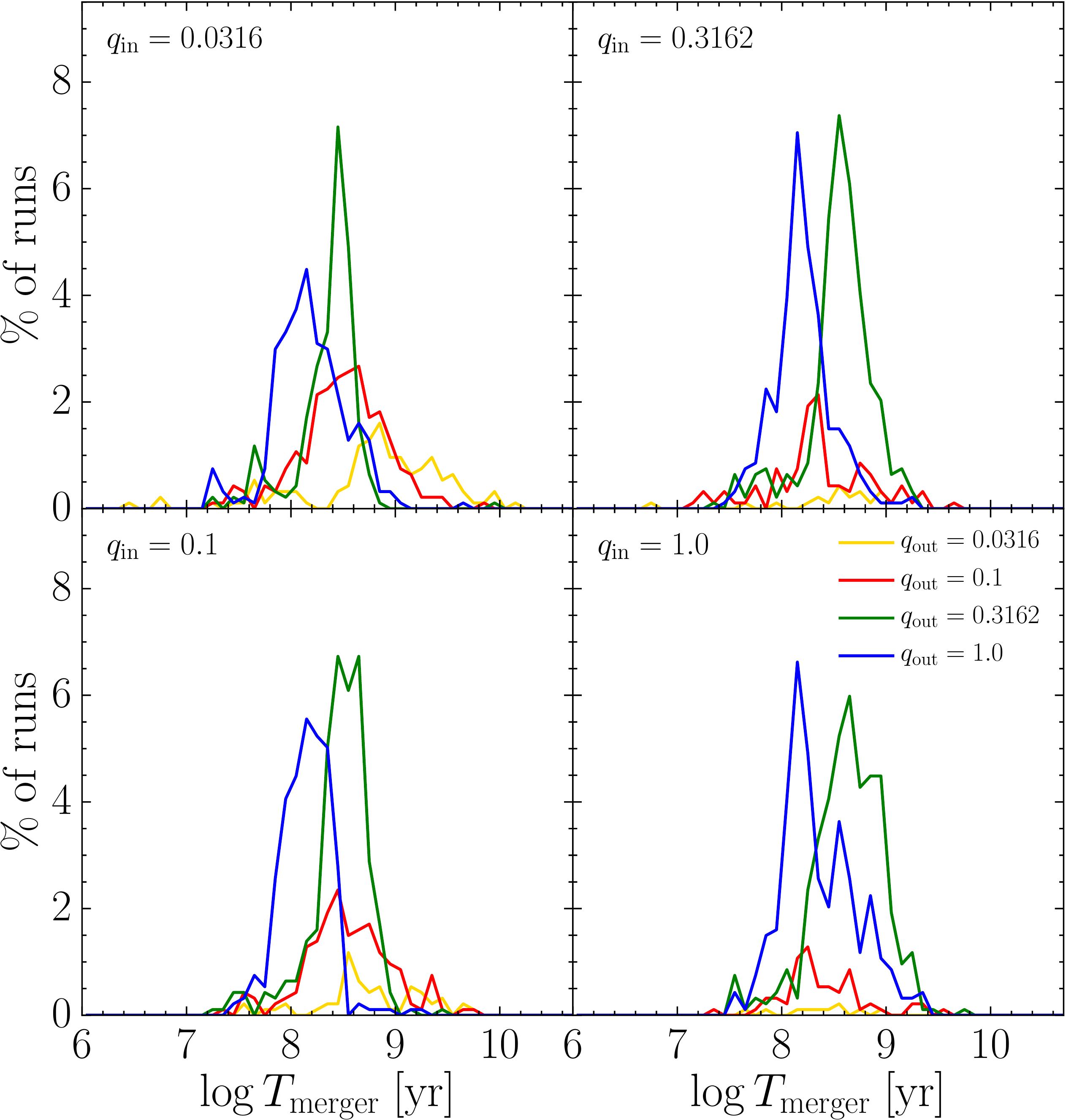}
	\caption{Distribution of merger-times, grouped according to the initial mass ratio of the inner (as labelled in the upper left corner of all panels) and outer binary (differentiated by colours as indicated in the lower right panel). The merger timescale only has a weak dependence on the outer mass ratio.}
	\label{fig:T_merger_mass_ratio}
\end{figure}
%%%%%%%%%%%%%%%%%%%%%%%%%%%%%%

The time spent by triplets before coalescence is shown in figure~\ref{fig:T_merger_all}. The merger-time distribution is remarkably well fit by a log-normal, with mean $\mu=8.4$ and standard deviation $\sigma=0.4$ in $\log(T/{\rm yr})$. The mean value $\mu=8.4$ corresponds to $\simeq 250$ Myr, i.e., a timescale substantially shorter than the Hubble time, indicating that the triplet channel can lead to fast mergers. Indeed, most of the time prior to merger is spent in the dynamical friction and stellar hardening dominated regimes, i.e., most of the time the intruder is far from the inner binary. Once a genuinely bound triplet is formed, secular and (in some cases) chaotic interactions drive the system to coalescence on a much shorter timescale. 

%%%%%%%%%%%%%%%%%%%%%%%%%%%%%%
\begin{figure*}
	\centering
	\includegraphics[scale=0.55]{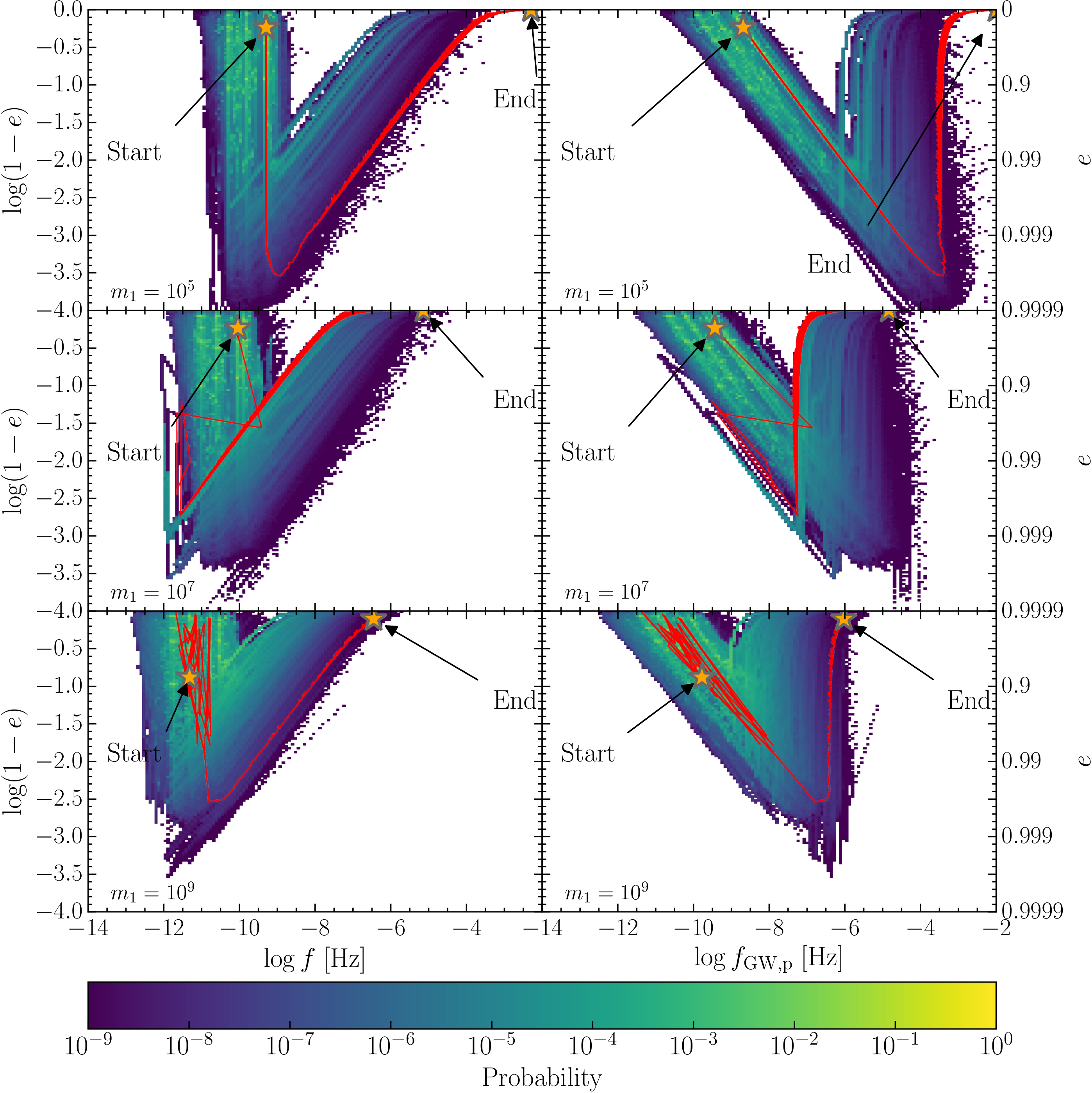}
	\caption{Evolution in the plane $(f,1-e)$ (left panels) and $(f_{\rm GW,p},1-e)$ (right panels) of all merging binaries for three selected values of $m_1= 10^5,10^7,10^9 \msun$ (from top to bottom). Colour-coded is the probability (see text for details about the computation) of finding a binary at given values of frequency and eccentricity. For each mass, superimposed in red is the evolutionary track of a representative binary that reached final coalescence. In the top panels such representative binary merges when the triplet is still in the hierarchical phase, while in the lower panels the binary experiences strong chaotic three-body interactions, clearly visible in the noisy change of orbital elements. The primary effects of the triple interactions (secular or chaotic) is the great increase of the orbital eccentricity, leading one of the pairs in the triplet to coalescence.}
	\label{fig:e_f_track}
\end{figure*}
%%%%%%%%%%%%%%%%%%%%%%%%%%%%%%

In figure~\ref{fig:T_merger_all} we also show, as a lower red histogram, the merger-time distribution of the ``post-ejection'' coalescences discussed in section 3.1. These events, which account for approximatively 1/5 of the total merger fraction and which are relatively more probable for high $m_1$ values (see table~\ref{tab:merger_summary}), involve an ejection, and a leftover inner binary that coalesces within a Hubble time under the effect of GW emission alone. The merger-time distribution is quite broad for these systems, and these MBHBs typically need a few Gyr to merge.  

In figure~\ref{fig:T_merger_mass_ratio} the merger-time distribution is shown for the different sampled values of 
$q_{\rm in}$  (in the four panels) and $q_{\rm out}$ (indicated by different colours in each panel). While there is no clear dependence of the merger 
timescale on $q_{\rm in}$, we note a weak dependence on $q_{\rm out}$, with $q_{\rm out}\simeq 1$ systems coalescing faster because of the stronger 
perturbations exerted by $m_3$ on the inner binary.

%%%%%%%%%%%%%%%%%%%%%%%%%%%%%%%%%%%%%%%%%%%%%%%%%%%%%%%%%%%%%%%%%%%%%%%%%%%%%%%%%%%%
\subsection{Eccentricity distribution}

%%%%%%%%%%%%%%%%%%%%%%%%%%%%%%
\begin{figure*}
 \begin{minipage}[c]{0.47\textwidth}
   \centering
   \includegraphics[scale=0.31]{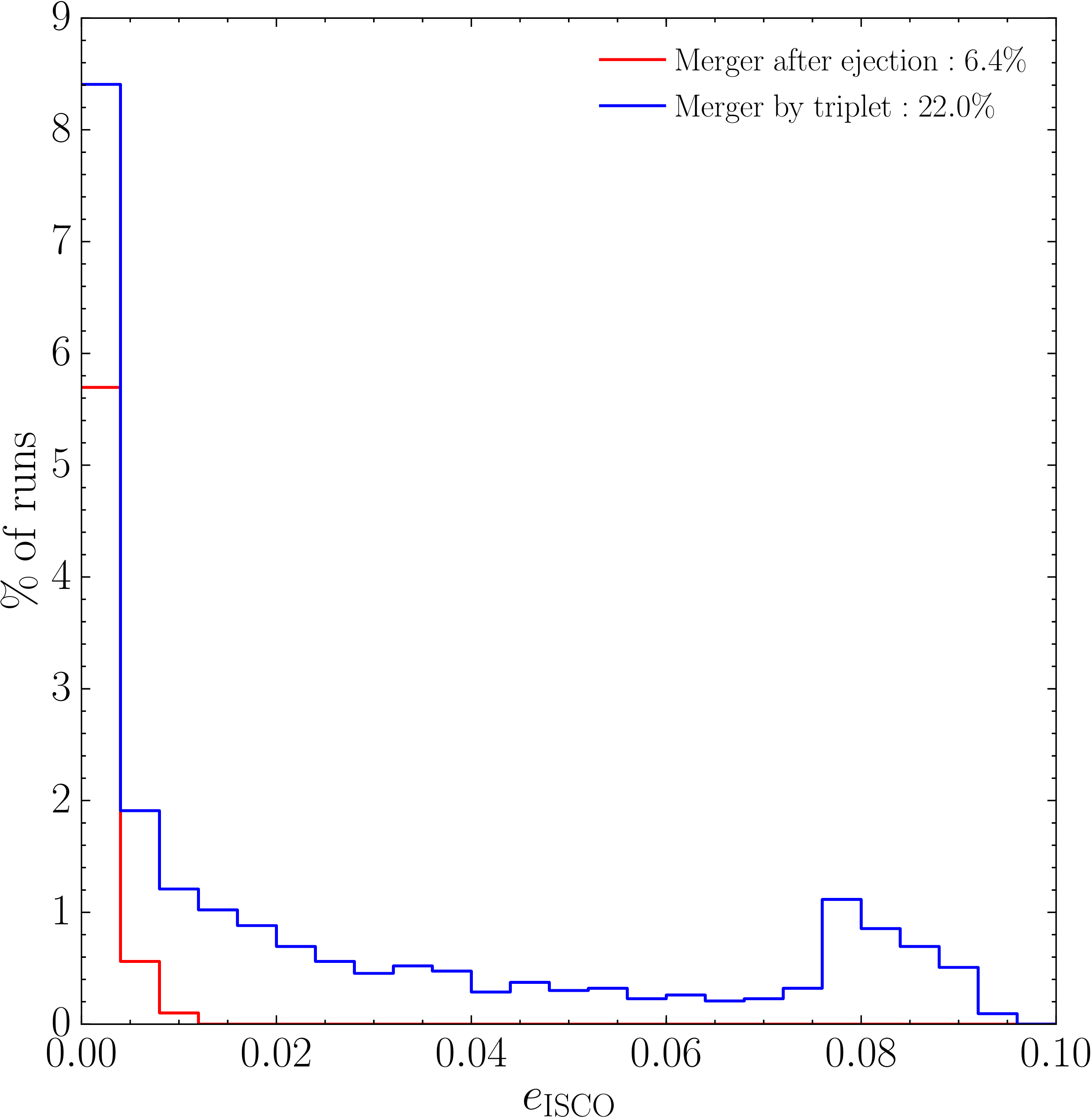}
 \end{minipage}
 \ \hspace{2mm} 
 \begin{minipage}[c]{0.47\textwidth}
  \centering
   \includegraphics[scale=0.31]{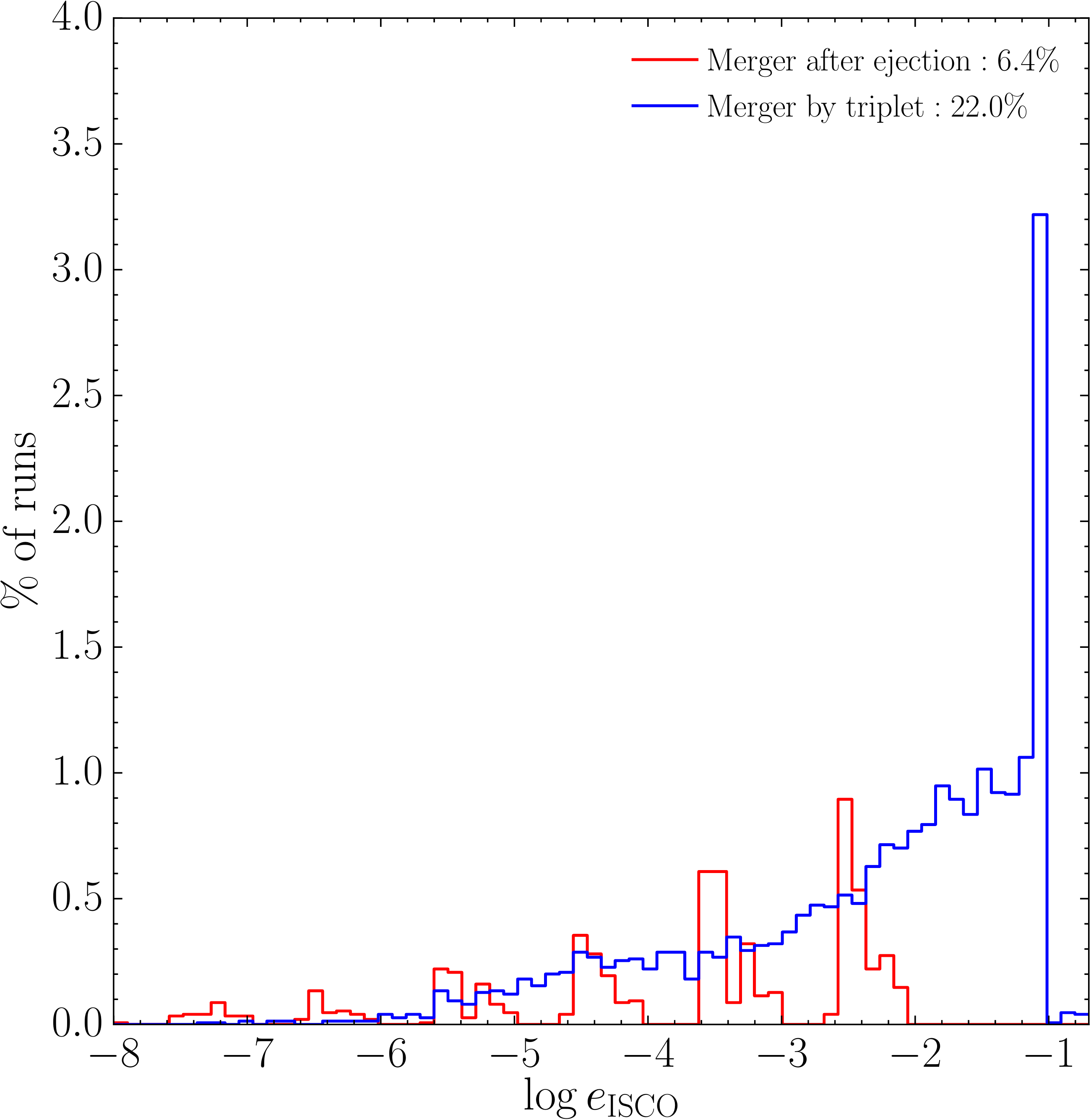}
 \end{minipage}
 \caption{Distribution of $e_{\rm ISCO}$ of all merging binaries. Colour code as in figure~\ref{fig:T_merger_all}. The binaries that merge after a strong interaction with the third body retain a larger eccentricity close to merger compared to those that are GW-driven. {\it Left panel}: linear scale. {\it Right panel}: log scale. Note the peculiar clustering of the red distribution in six different blocks corresponding to the six different values of $m_1$, i.e., from, left to right, $10^5\msun$ to $10^{10}\msun$.} 
  \label{fig:e_merger}
\end{figure*}
%%%%%%%%%%%%%%%%%%%%%%%%%%%%%%

Of particular importance for GW emission from MBHBs in a cosmological setting is the study of the eccentricity evolution of merging binaries. 
In the left panels of figure~\ref{fig:e_f_track} we track the evolution of the merging binaries in the orbital frequency vs circularity plane $(f,1-e)$, colour coding the probability of finding a binary at given values of eccentricity and frequency. We first discretise the 
$(\log f,\log (1-e))$ space in the range $-14\leq \log f \leq -2$ and $-4 \leq \log(1-e) \leq 0$ on a $150\times 150$ grid equally spaced along each direction. 
Then, we evaluate the time spent by each merging binary in each of the 22,500 bins of the grid during its evolution, then sum  over all merging binaries,    and normalise to the total time spent over all bins by all binaries. In this way, we obtain a bivariate normalised function that gives the probability of finding a binary in a given logarithmic two-dimensional interval of frequency and circularity. We construct this function for the six sampled values of $m_1$, and show in figure~\ref{fig:e_f_track}, left panels, three cases ($m_1=10^5,10^7,10^9$ M$_\odot$). The evolutionary tracks of single illustrative merging binaries are shown as a red line.

In the orbital frequency-eccentricity plane, a typical stalled inner binary starts its evolution in the upper left corner, i.e., at large separation (i.e., low $f$) and with an eccentricity given by one of the 4 values of $e_{\rm in}$ that we sample 
(see table~\ref{tab:param_space}). As soon as the perturbations due to the approaching third body become significant, the inner binary becomes more eccentric. If the system undergoes K-L resonances, the eccentricity actually oscillates on the K-L timescale between high and low values (these oscillations are not visible in
figure~\ref{fig:e_f_track} due to the scale used), with a secular shift to higher values because of the perturbation exerted by an increasingly closer $m_3$. The orbital frequency (i.e., the separation) of the inner binary stays nearly constant during this evolutionary phase. An example is given by the red line shown in the $m_1=10^5$ M$_\odot$ case (figure~\ref{fig:e_f_track}, upper left panel). 
When chaotic interactions are instead the main driver of the binary evolution, $f$ can show large, random variations, as exemplified by the tracks in the middle and lower left panels of figure~\ref{fig:e_f_track}. 

In any case, when the eccentricity becomes very high, $\gsim 0.99$, GW emissions starts dominating the dynamics, increasing the orbital frequency and circularising the orbit until coalescence, as can be seen from the rising branch of the red tracks. The colour code shows that this circularisation phase is much shorter than the 
preceding evolution. The maximum eccentricity reached (the turnover point in the evolutionary tracks) mainly depends on the mass of the inner binary, i.e.  the lower the mass, the higher the maximum eccentricity. In fact, for more massive binaries GWs start dominating sooner during the evolution, hence determining the earlier orbital circularisation. This behaviour is clearly visible in figure~\ref{fig:e_f_track}, moving from the top panel ($m_1=10^5$ M$_\odot$) to the bottom one ($m_1=10^9$ M$_\odot$). Note that for more massive systems the orbital frequency at merger is necessarily lower, since it scales as $M^{-1}$, 
where $M\equiv m_1+m_2$.

It is of a certain interest to analyse the same evolution not in terms of the  orbital frequency, but rather in terms of the peak frequency of the GW power spectrum \citep{Wen2003}

%%%%%%%%%%%%%%%%
\begin{equation}
f_{\rm GW,p} = \dfrac{1}{\pi}\sqrt{\dfrac{G M}{[a(1-e^2)]^3}}(1+e)^{1.1954}, 
\end{equation}
%%%%%%%%%%%%%%%%%
which is clearly larger for more eccentric binaries. (Note that this equation essentially means that GWs are mainly emitted at the pericentre passages). The probability distribution in the $(\log f_{\rm GW,p},\log(1-e))$ plane is shown in the right panels of figure~\ref{fig:e_f_track}, for the same three values of $m_1$ considered before. 

During the initial phase of eccentricity growth, irrespective of the evolution driver (K-L resonances or chaotic interactions), 
the orbital frequency does not change (left panels in figure~\ref{fig:e_f_track}), but $f_{\rm GW,p}$ increases because of its dependence on $e$. As soon as 
GWs take over, the orbit circularises fast while maintaining an almost constant $f_{\rm GW,p}$ until the very last phase of the evolution. This means that during the circularisation phase, the binaries maintain a fixed pericentre separation while their the semi-major axis shrinks. Once the circularisation is completed, 
GW losses keep shrinking the semi-major axis. Therefore, $f_{\rm GW,p}$ increases and eventually becomes twice the orbital frequency ($f_{\rm GW,p} = 2 f$), as expected for circular binaries at leading PN order. Like the orbital frequency, in the GW dominated regime $f_{\rm GW,p}$ is lower for larger masses.

A particularly interesting result of our simulations is that
more massive binaries
merge with a slightly larger eccentricity compared to their low mass counterparts,
 despite their maximum eccentricity being comparatively lower. This is essentially due to the shorter timescale of the inspiral phase, which results in a sizeable residual eccentricity, 
as can be seen in figure~\ref{fig:e_f_track} by observing the positions marked as ``End''. 

Figure~\ref{fig:e_merger} shows the distribution of the eccentricity of merging MBHBs at an arbitrary spatial threshold corresponding to the last stable circular orbit of a non-spinning MBH with mass equal to the total binary mass, i.e., $r_{\rm ISCO}=6R_G$ with $R_G=GM/c^2$. In the left panel, we plot the distribution on a linear scale, while in the right panel we show the logarithmic version of the same distribution. It is remarkable that  the distribution extends up to $e_{\rm ISCO} \simeq 0.1$, as shown by the blue histogram. 

%%%%%%%%%%%%%%%%%%%%%%%%%%%%%%
\begin{figure}
	\centering
	\includegraphics[scale=0.31]{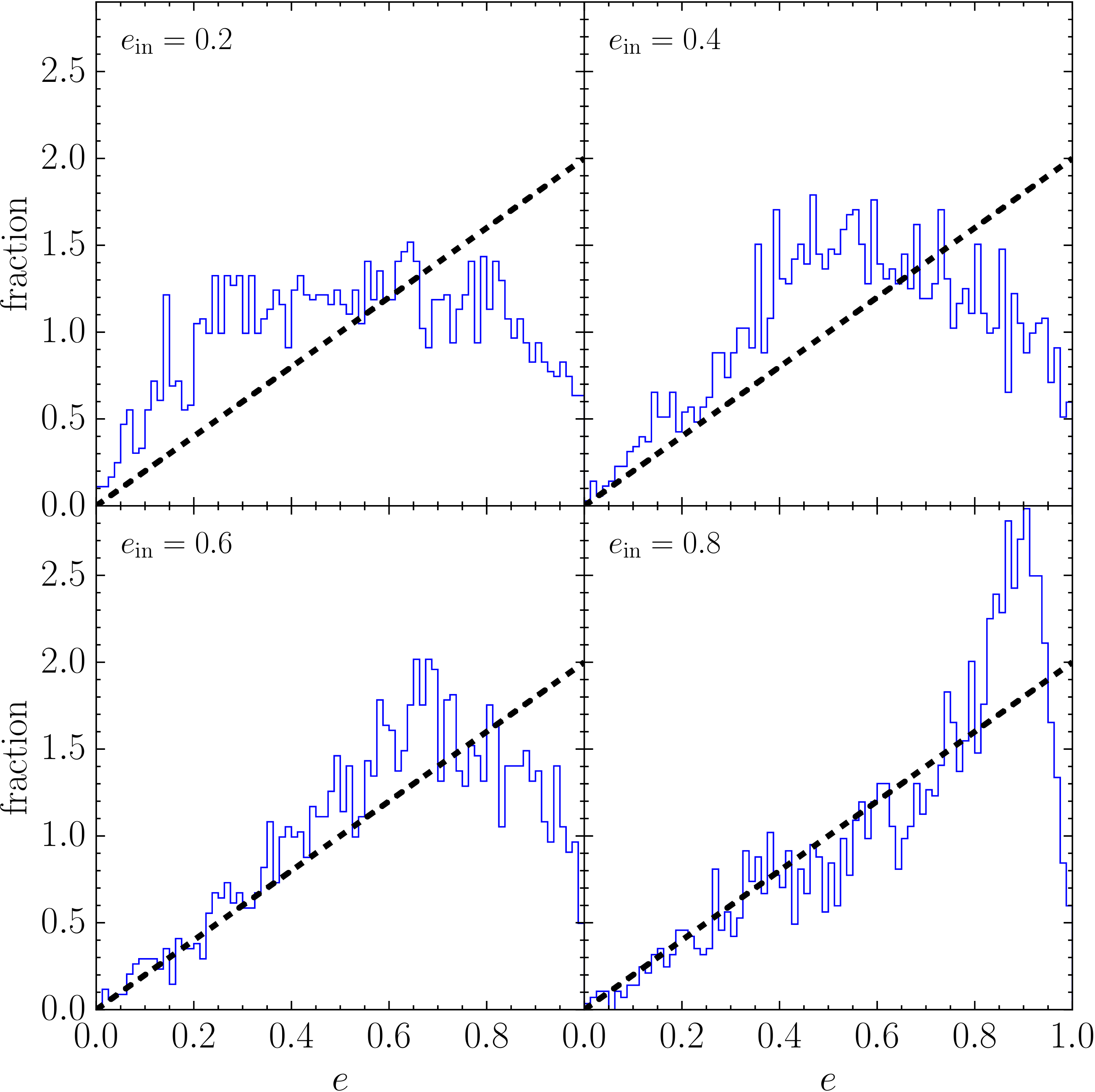}
	\caption{Eccentricity distribution according to the initial value of $e_{\rm in}$, for those binaries that, after the ejection of one of the MBHs, {\it do not} merge within a Hubble time. The eccentricity approximately follows a thermal distribution (i.e., $f(e) = 2 e$, represented as a black dashed line in the figure) in the range $0\lsim e \lsim e_{\rm in}$. At higher eccentricities the distribution shows a turnover, as some binaries are driven to coalescence (and therefore counted as mergers) by GW emission.}
	\label{fig:e_ejection_slice}
\end{figure}
%%%%%%%%%%%%%%%%%%%%%%%%%%%%%%

%%%%%%%%%%%%%%%%%%%%%%%%%%%%%%
\begin{figure}
	\centering
	\includegraphics[scale=0.31]{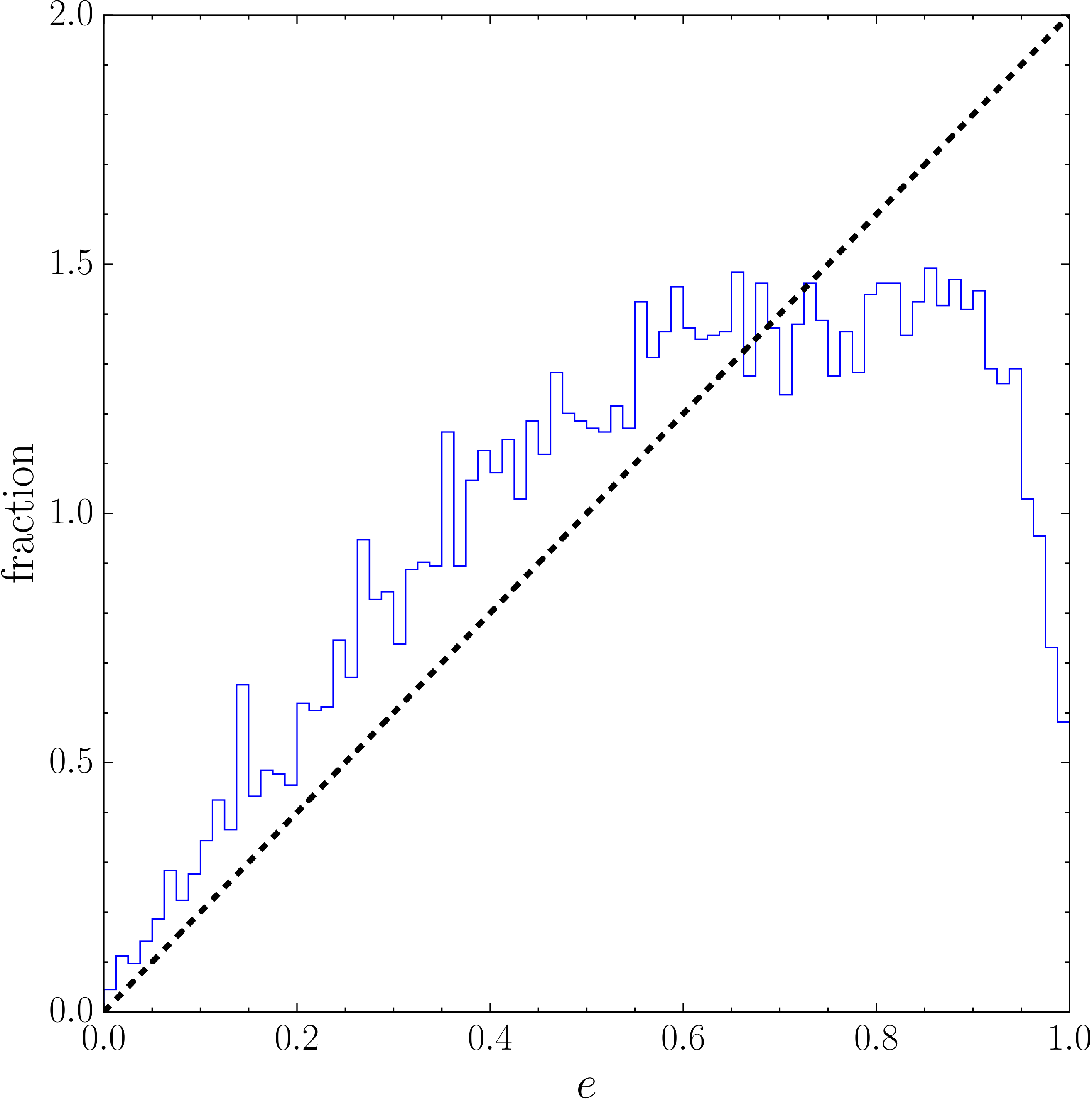}
	\caption{Same as figure~\ref{fig:e_ejection_slice}, summing over all $e_{\rm in}$. Note that the slope is steeper than thermal because of the contribution of low $e_{\rm in}$.}
	\label{fig:e_ejection}
\end{figure}
%%%%%%%%%%%%%%%%%%%%%%%%%%%%%%

In the same figure, the red histogram refers to the  ``post-ejection'' coalescences, which we recall account for $\simeq 1/5$ of the total. As expected, given that the final coalescence is purely driven by GW emission, these mergers are much less eccentric. Their low residual eccentricity has a marked dependence of the mass of the triplet, as can 
be inferred from the logarithmic version of the plot. Indeed, the right panel of figure~\ref{fig:e_merger} shows that the $e_{\rm ISCO}$ distribution clusters around 6 different values, corresponding to the 6 values of $m_1$ that we have sampled, with $m_1$ increasing from left to right. Note that the distribution of $e_{\rm ISCO}$ for the mergers driven by triple interactions does not show any clustering, as expected in the case of coalescences induced by dynamical processes.  

Finally, in figure~\ref{fig:e_ejection_slice}, for different values of $e_{\rm in}$, we plot the eccentricity distribution of the inner binaries that {\it did not} merge within a Hubble time after the ejection of one of the three bodies. The distribution is approximatively thermal \citep[i.e., $p(e)\propto e$, see, e.g.,][for further details]{Jeans1919,Heggie1975} in the range from $0\lsim  e \lsim e_{\rm in}$. This behaviour is typical of binaries that have experienced strong dynamical encounters during their evolution. The distribution shows a turnover at $e\gsim e_{\rm in}$, due to the fact that binaries with a higher eccentricity merge within a Hubble time, and are therefore not counted in the shown distribution.  
In figure~\ref{fig:e_ejection} the same distribution is plotted summing over all $e_{\rm in}$. In this case, the slope results steeper than thermal because of the piling of binaries with low $e_{\rm in}$, which obey a thermal distribution only in a narrower range of $e$.

%%%%%%%%%%%%%%%%%%%%%%%%%%%%%%%%%%%%%%%%%%%%%%%%%%%%%%%%%%%%%%%%%%%%%%%%%%%%%%%%%%%%
%%%%%%%%%%%%%%%%%%%%%%%%%%%%%%%%%%%%%%%%%%%%%%%%%%%%%%%%%%%%%%%%%%%%%%%%%%%%%%%%%%%%
\section{Discussion}
\label{sec:discusssion}

%%%%%%%%%%%%%%%%%%%%%%%%%%%%%%%%%%%%%%%%%%%%%%%%%%%%%%%%%%%%%%%%%%%%%%%%%%%%%%%%%%%%
\subsection{Implications for the emission of gravitational waves}
The results presented in the previous section suggest that MBH triplets might have a critical impact on the emission and detection of low-frequency gravitational waves. Although two forthcoming papers in this series are devoted to derive and analyse in depth the implications for LISA and PTAs separately, we preview here some relevant points.

The fact that about 30\% of triple systems lead to coalescence of a MBHB implies that this is an effective ``last resort'' to overcome the final-parsec problem, should all other dynamical mechanisms fail. If the average galaxy undergoes a fairly large number of mergers during its cosmic history, then triple MBH interactions guarantee that a significant fraction of these galaxy mergers leads to a MBHB coalescence. \citep[See, also,][]{Mikkola1990,Heinamaki2001,Blaes2002,Hoffman2007,Kulkarni2012}. For example, under the simplifying assumption that the parameters of the forming triplets are uniformly distributed in the range explored in this study, this fraction is expected to be around 30\%. Therefore, compared to a scenario where MBHBs merge efficiently, the merger rate should be at most suppressed by a factor of $\simeq 3$. This is particular encouraging for low-frequency GW probes. Indeed,  even if MBHB stalling turns out to be a problem, LISA detection rates would be affected by a factor $\simeq 3$ only, while the stochastic GW background in the PTA band would only be suppressed by a factor $\sqrt{3}$ (since the GW background is proportional to the square root of the number of mergers). Conversely, if the average galaxy undergoes $\lsim 1$ merger during its cosmic history, MBH triplets would not form frequently. In this scenario, MBHB stalling would result in a severe suppression of any low-frequency GW signals, posing a potential threat to PTAs and LISA. In accompanying and forthcoming papers of the series, we will couple our library of simulations to a semianalytic model for galaxy and MBH evolution, to explore which type of scenarios is more likely to occur in Nature, and to properly quantify the fraction of galaxy mergers resulting in MBHB coalescences.

%%%%%%%%%%%%%%%%%%%%%%%%%%%%%%%%%%%
\begin{figure}
	\centering
	\includegraphics[scale=0.31]{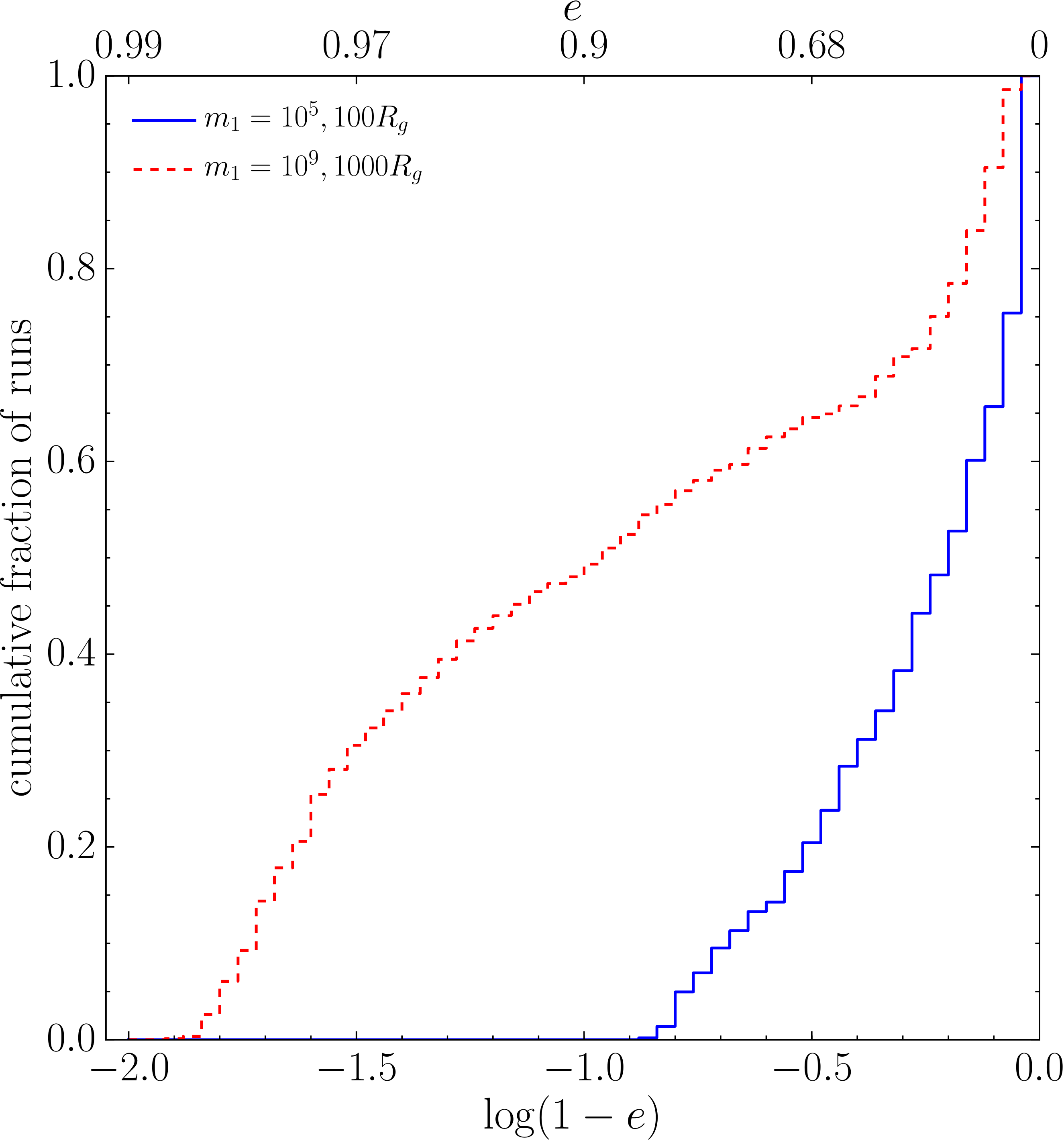}
	\caption{Cumulative distribution of $1-e$ of all merging binaries with $m_1 = 10^5\msun$ (blue solid line) and $m_1 = 10^9\msun$ (red dashed line). In the case of $m_1=10^5\msun$ ($10^9\msun$) the distribution is evaluated when the separation is $100$ ($1000$) $R_G$, most relevant for LISA (PTAs).}
\label{fig_ecc_LISA_PTA}        
\end{figure}
%%%%%%%%%%%%%%%%%%%%%%%%%%%%%%%%%%%

Triple interactions also leave a distinctive imprint in the eccentricity distribution of merging MBHBs. In fact, whether secular processes or chaotic dynamics dominate the evolution, the coalescence is triggered when one of the MBH pairs is eccentric enough that a significant amount of GWs is emitted at subsequent pericentre passages. The net result is that triplet-induced MBHB coalescences typically involve eccentric systems. Even at separation $\sim r_{\rm ISCO}$, eccentricity can still be as high as 0.1, and is $\gsim 0.01$ for more than $50\%$ of the binaries. 

LISA is mostly sensitive to $10^5\msun-10^6\msun$ MBHBs throughout the Universe. Those systems typically enter the detector band at separations around $100R_G$. The cumulative eccentricity distribution of merging systems for all simulations with  $m_1=10^5\msun$ is shown in figure \ref{fig_ecc_LISA_PTA}. Although skewed towards $e=0$, the distribution extends to $e\approx 0.8$, with about 30\% of the systems having $e\gsim 0.5$. Therefore, high eccentricities in the LISA band might be the smoking gun of triple-driven coalescences, and waveforms accurate up to high eccentricities might be necessary for proper recovery of the source parameters. Conversely, PTAs are sensitive to masses $\gsim 10^8\msun$ at low redshift. In figure \ref{fig_ecc_LISA_PTA} we also show the eccentricity distribution of merging systems for all simulations with  $m_1=10^9\msun$ at a separation of $1000R_G$, which are representative of the sources dominating the GW signal in the nHz band. Note that the distribution extends to $e\approx 0.99$, and about $50\%$ of the systems have eccentricity in excess of $0.9$. Thus, in a Universe dominated by triple interactions, the PTA signal is expected to be dominated by very eccentric binaries. 

A further consequence of high eccentricities is the possibility of generating bursts of GWs. In practice, binaries with high $e$ mostly emit GWs at every pericentre passage, resulting in a ``burst signal''  well localised in time and spread (in frequency) over a large number of harmonics~\citep{Amaro-Seoane2010}. As an example, massive binaries with orbital periods of hundreds of years can emit month-long bursts in the PTA frequency band, while lighter binaries with periods of several months can emit bursts detectable by LISA. This latter case is particularly interesting, because it might enhance the number of LISA detections well beyond the nominal MBHB merger rate. We will investigate this possibility in a forthcoming paper.

%%%%%%%%%%%%%%%%%%%%%%%%%%%%%%%%%%%%%%%%%%%%%%%%%%%%%%%%%%%%%%%%%%%%%%%%%%%%%%%%%%%%
\subsection{Effect of massive intruders}
\begin{figure}
%\hspace{-0.7cm}
	\centering
	\includegraphics[scale=0.33]{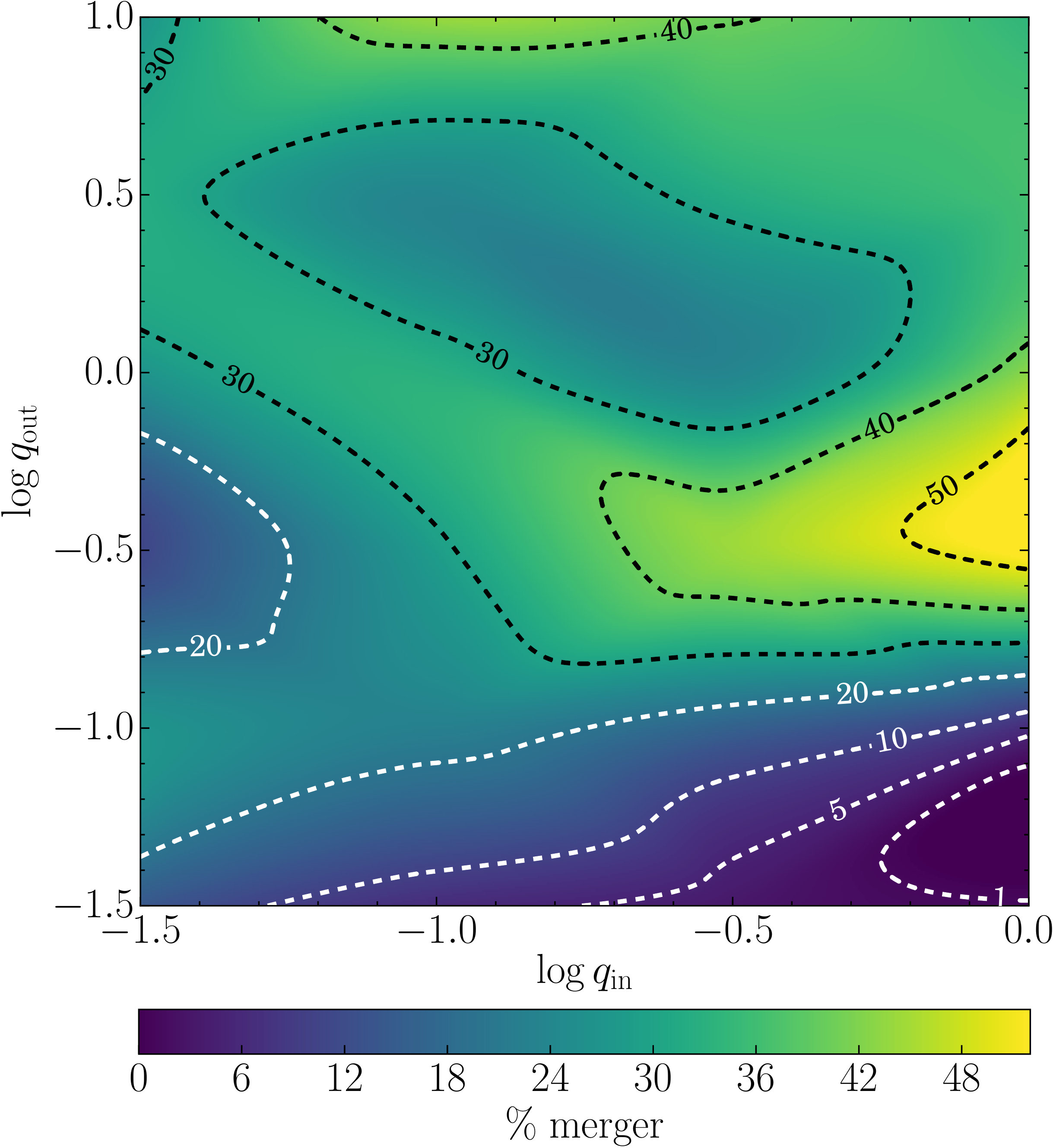}
	\caption{Same as figure \ref{fig:merger_fraction_mass_ratio}, but now extended up to $q_{\rm out}=10$. Only the case 
	$m_1=10^9\msun$ is considered.}
\label{fig_qin_qout_extended}        
\end{figure}

Although the inner mass ratio is $q_{\rm in}\leq1$ by definition, the outer mass ratio $q_{\rm out}=m_3/(m_1+m_2)$ may be larger than one if the intruder is more massive than the bound binary. This can be relevant in a hierarchical structure-formation scenario, where a pre-existing stalled binary can be involved in a merger with a third, more massive black hole brought on by a galaxy merger. To explore the possible outcomes of this kind of configurations, we have run an additional set of simulations for the case $m_1=10^9$ with $q_{\rm out}=3,\,10$. Although in this case one might expect the stellar potential to be dominated by the host galaxy of the intruder $m_3$, we ignore this fact and simply centre the stellar potential in the (initial) centre of mass of the bound inner binary. This should not significantly affect the outcome of the simulation, at least qualitatively and in a statistical sense, since we find that the stellar potential has little effect on both the secular K-L evolution and the later chaotic phase (if present). 

Results are shown in figure \ref{fig_qin_qout_extended}, where the original parameter space of figure \ref{fig:merger_fraction_mass_ratio_bin} is extended up to $q_{\rm out}=10$. As one might expect, the merger fraction stays quite high, around 30\% when $q_{\rm out}\gsim 1$. This is the result of two competing effects. On the one hand the K-L timescale is inversely proportional to $m_3/(m_1+m_2)$; a massive intruder can therefore excite the eccentricity of the inner binary several times, favouring its prompt coalescence. On the other hand, at inclinations where K-L resonances are not very efficient, the unequal mass ratio favours the ejection of the lightest black hole, suppressing the fraction of systems that can merge in chaotic interactions. In any case, it is therefore important to take into account this kind of configurations when coupling libraries of triplet outcomes to semianalytic galaxy-formation models, because they can have an impact on the global merger rate MBHBs.

%%%%%%%%%%%%%%%%%%%%%%%%%%%%%%%%%%%%%%%%%%%%%%%%%%%%%%%%%%%%%%%%%%%%%%%%%%%%%%%%%%%%
\subsection{Comparison with previous work}

We compare the results of the analysis of our simulations with the work of \citet{Hoffman2007}, 
which our investigations is similar in spirit, although we introduce some important novelties. The major differences are, i) the surveyed parameter space, which we extend to a wider range of masses, mass ratios and initial eccentricities, ii) our conservative choice of not considering a DM component and, iii) the introduction of the conservative 1PN dynamics.

By comparing the merger fractions (their table~1 compared to our table~\ref{tab:merger_summary}), we immediately note that their values are typically a factor $\simeq 3$ larger than our results, even if ``post-ejection'' merger binaries are added to our total. The discrepancy can be understood by analysing the differences in the two implementations:

%%%%%%%%%%%%%%%%%%%
\begin{itemize}
\item All the simulations in \citet{Hoffman2007} have nearly the same total mass ($\simeq 6\times 10^8 \msun$), and moreover they considered systems, except in one case, in which the MBHs are nearly equal-mass. The interaction of nearly equal-mass objects produces the highest merger fraction, as can be seen from, e.g., our table~\ref{tab:cfr_PN}, where in the nearly equal-mass case the merger fraction is $\simeq 50\%$. In the only case in which \citet{Hoffman2007} consider a lower mass ratio, the  merger fraction decreases below $\lsim 70\%$, suggesting that the chosen mass ratio is in fact one of the reasons of the higher  merger percentage. In practice, when restricting to comparable mass ratios, the merger fractions differ by less than a factor $\simeq 2$ ($\approx 50\%$ vs $\approx 85\%$).

\item The 1PN dynamics can further contribute to the difference. Even though it is mostly effective at low mass ratios (as can be appreciated from figure~\ref{fig:cfr_w_wo_PN}), the merger fraction can still be about 10\% higher even at comparable mass ratios when the 1PN relativistic precession is neglected.

\item The absence of a DM halo and the conservative threshold for MBH ejection assumed in our simulations also plays a role. As already discussed in Section \ref{sec:ICs}, our choice implies that a larger number of MBHs are ejected, compared to \citet{Hoffman2007}. This is confirmed by their run featuring a less massive DM halo, where the merger fraction drops to $\sim 70\%$. As already discussed, our choice is meant to be conservative, in the sense that outside the stellar bulge significant triaxiality and asymmetries in the potential will likely prevent the ejected MBH from returning back to the centre \citep{Guedes2009,Sijacki2011}. We will explore this issue in greater detail in a future work. 
\end{itemize} 
%%%%%%%%%%%%%%%%%%%%%%%%%

In summary, we argue that the combination of the three points above fully explains the differences between the two studies. 
   
%%%%%%%%%%%%%%%%%%%%%%%%%%%%%%%%%%%%%%%%%%%%%%%%%%%%%%%%%%%%%%%%%%%%%%%%%%%%%%%%%%%%
%%%%%%%%%%%%%%%%%%%%%%%%%%%%%%%%%%%%%%%%%%%%%%%%%%%%%%%%%%%%%%%%%%%%%%%%%%%%%%%%%%%%
\section{Conclusions}
\label{sec:conclusions}

In the present study we have utilised the three-body integrator presented in \citet{Bonetti2016} to investigate the outcome of MBH triple interactions over a vast parameter space. The code evolves the dynamics of MBH triplets including a variety of relevant factors, such as an external galactic potential, the dynamical friction against the stellar background, the stellar hardening, and the PN corrections to the equations of motion consistently derived from the three-body PN Hamiltonian. The set-up of the code is tuned to capture the physics relevant to three-body interactions of MBHs originated by repeated galaxy mergers in hierarchical cosmologies. We have explored the parameter space relevant to this specific context by considering primary MBH masses in the range $10^5\msun\leq m_1\leq 10^{10}\msun$, a variety of inner and outer binary mass ratios in the range $0.03-1$, and several inner and outer binary initial eccentricities and mutual orbit inclinations as detailed in table \ref{tab:param_space}.  We have integrated a grand total of 14,976 configurations with the goal of deriving the fraction of merging systems as a function of the relevant parameters, as well as the typical merger timescales and properties of the coalescing MBHs.

Our main results can be summarised as follows:

%%%%%%%%%%%%%%%%%%%%%%%
\begin{itemize}
\item The fraction of systems experiencing the merger of one of any pair of MBHs in the triplet is about 30\%. About 4/5 of these mergers are promptly induced during the three-body interaction, whereas about 1/5 are driven by GW emission following the ejection of the lightest black hole.
\item Prompt mergers are induced both by secular K-L evolution and chaotic dynamics. The former is more efficient for massive intruders (large $q_{\rm out}$) and eccentric inner binaries, while the latter is most efficient for nearly equal-mass systems. Overall, the merger fraction is higher for large $q_{\rm out}$, reaching 40\% of the systems.
\item The 1PN terms in the equations of motion are important at low $q_{\rm in}/q_{\rm out}$. Neglecting such correction, the fraction of low-$q$ systems leading to a merger goes from $\approx 20\%$ to about $40\%$. This happens because 1PN precession destroys the K-L resonances, preventing the inner binary from reaching eccentricities high enough to merge under the effect of GW backreaction.
\item The typical timescale for prompt mergers is well described by a log-normal distribution centred around $\log(T/\rm yr)=8.4$ with a dispersion of 0.4 dex, almost independent of the masses and mass ratios involved. Note that this timescale is dominated the orbital decay of the intruder, driven by dynamical friction and stellar hardening. Once the chaotic phase starts, mergers can be triggered within a few Myr Mergers following a MBH ejection occur on much longer timescales, of the order of several Gyr
\item Merging binaries are generally driven to eccentricities in excess of 0.9 and up to 0.9999 in some cases, and even at coalescence binaries can retain a significant eccentricity, up to 0.1. Binaries driven by triple MBH interactions can therefore have eccentricities in excess of 0.5 when entering the LISA band, and in excess of 0.9 in the PTA frequency range. 
\end{itemize}
%%%%%%%%%%%%%%%%%%%%%%%

Compared to the merger fractions found in previous studies \citep{Hoffman2007}, our numbers are significantly lower. This is because our parameter space includes lower mass ratios (which give a lower merger fraction), PN dynamics (which partially suppresses K-L resonances) and a conservative prescription for ejection (which does not include the galactic DM potential). Essentially, we consider a MBH as ejected forever once it reaches the outskirts of the stellar bulge. This is justified by the fact that triaxiality and potential asymmetries will likely prevent the MBH from sinking back to the centre in less than a few Gyr. In a future work, we plan to include a triaxial DM halo in our simulations, and study the impact on the MBHB merger fraction.

The aforementioned results indicate that MBH triplets can have a significant impact on MBH evolution and on future GW detections by LISA and PTAs. In particular, three-body interactions provide a partial solution to the final-parsec problem, even if every all other binary shrinking mechanisms fail. This should guarantee a fairly large number of detectable GW sources, both by  LISA and PTAs, regardless of the details of the interaction between MBHBs and their environment in galactic nuclei. In two papers, one accompanying and one forthcoming, we investigate this issue further by coupling our extensive library of MBH-triplet simulations to a state-of-the-art semianalytic model of galaxy and MBH evolution. 

%%%%%%%%%%%%%%%%%%%%%%%%%%%%%%%%%%%%%%%%%%%%%%%%%%%%%%%%%%%%%%%%%%%%%%%%%%%%%%%%%%%%
%%%%%%%%%%%%%%%%%%%%%%%%%%%%%%%%%%%%%%%%%%%%%%%%%%%%%%%%%%%%%%%%%%%%%%%%%%%%%%%%%%%%
\section*{Acknowledgements}
MB and FH acknowledge partial financial support from the INFN TEONGRAV specific initiative. MB acknowledge the CINECA award under the ISCRA initiative, for the availability of high performance computing resources and support. This work was supported by the H2020-MSCA-RISE-2015 Grant No. StronGrHEP-690904. AS is supported by a University Research Fellowship of the Royal Society. This work has made use of the Horizon Cluster, hosted by the Institut d'Astrophysique de Paris. We thank Stephane Rouberol for running smoothly this cluster for us.

%%%%%%%%%%%%%%%%%%%%%%%%%%%%%%%%%%%%%%%%%%%%%%%%%%%%%%%%%%%%%%%%%%%%%%%%%%%%%%%%%%%%
\bibliographystyle{mn2e}
\bibliography{Bibliography} 
%%%%%%%%%%%%%%%%%%%%%%%%%%%%%%%%%%%%%%%%%%%%%%%%%%%%%%%%%%%%%%%%%%%%%%%%%%%%%%%%%%%%
%%%%%%%%%%%%%%%%%%%%%%%%%%%%%%%%%%%%%%%%%%%%%%%%%%%%%%%%%%%%%%%%%%%%%%%%%%%%%%%%%%%%
\clearpage
\appendix
%%%%%%%%%%%%%%%%%%%%%%%%%%%%%%%%%%%%%%%%%%%%%%%%%%%%%%%%%%%%%%%%%%%%%%%%%%%%%%%%%%%%
\section{Merger fractions}
\label{sec:app_A}
In this appendix we report the tables of the merger fraction per range of $m_1$ sliced according to various IC parameters, i.e., mass ratios, outer eccentricity and initial relative inclination. 

\clearpage
%%%%%%%%%%%%%%%%%%%%%%%%%%%%%%%%%%%%%%%%%%%%%%%%%%%%%%%%%%%%%%%%%%%%%%%%%%%%%
%%%%%%%%%%%%%%%%%%%%%%%%%%%%%%%%%%%%%%%%%%%%%%%%%%%%%%%%%%%%%%%%%%%%%%%%%%%%%
%m1 = 1e10

\begin{table}
\caption{Results $m_1=10^{10}\rm M_{\odot}$}
\label{tab:m1_10}
\begin{tabular}{lcccl}
\hline
$m_1 = 10^{10} \ \rm M_{\odot}$ & \multicolumn{4}{c}{\% Mergers} \\
$q_{\rm in}/q_{\rm out}$	 & $m_1$-$m_2$  & $m_1$-$m_3$  & $m_2$-$m_3$ & Total \\
\hline
0.0316/0.0316 &  21.2 &   9.6 &   0.0 &   30.8(32.7) \\
0.0316/   0.1 &  33.3 &   6.4 &   0.0 &   39.7(11.5) \\
0.0316/0.3162 &  19.9 &   4.5 &   0.0 &   24.4(7.7 ) \\
0.0316/     1 &  32.1 &   1.3 &   5.8 &   39.1(7.7 ) \\
   0.1/0.0316 &   7.7 &   7.7 &   0.0 &   15.4(15.4) \\
   0.1/   0.1 &  21.8 &  13.5 &   1.9 &   37.2(23.1) \\
   0.1/0.3162 &  36.5 &   8.3 &   3.2 &   48.1(10.3) \\
   0.1/     1 &  18.6 &   4.5 &   4.5 &   27.6(7.1 ) \\
0.3162/0.0316 &   4.5 &   1.3 &   0.0 &    5.8(10.9) \\
0.3162/   0.1 &  16.7 &   3.8 &   1.3 &   21.8(12.2) \\
0.3162/0.3162 &  37.2 &  14.7 &   1.9 &   53.8(15.4) \\
0.3162/     1 &  19.2 &   7.7 &   7.1 &   34.0(7.1 ) \\
     1/0.0316 &   0.6 &   1.3 &   0.6 &    2.6(5.1 ) \\
     1/   0.1 &   9.6 &   1.3 &   0.0 &   10.9(12.2) \\
     1/0.3162 &  32.7 &  16.7 &  10.9 &   60.3(9.6 ) \\
     1/     1 &  25.6 &  18.6 &  15.4 &   59.6(14.7) \\
\hline
Average		  &  21.1 &   7.6 &   3.3 &   31.9(12.7) \\
\hline
\end{tabular}
\end{table}

\begin{table}
\begin{tabular}{lcccl}
\hline
$m_1 = 10^{10} \ \rm M_{\odot}$ & \multicolumn{4}{c}{\% Mergers} \\
$e_{\rm in}/e_{\rm out}$ 	 & $m_1$-$m_2$  & $m_1$-$m_3$  & $m_2$-$m_3$ & Total \\
\hline
0.2/0.3 & 18.3 & 6.7  & 5.8 & 30.8(18.3) \\
0.2/0.6 & 8.2  & 9.6  & 2.9 & 20.7(6.7 ) \\
0.2/0.9 & 8.2  & 13.9 & 3.8 & 26.0(3.8 ) \\
0.4/0.3 & 24.5 & 4.8  & 3.4 & 32.7(7.2 ) \\
0.4/0.6 & 15.9 & 4.8  & 2.9 & 23.6(14.9) \\
0.4/0.9 & 13.9 & 11.5 & 2.9 & 28.4(7.2 ) \\
0.6/0.3 & 28.8 & 8.2  & 1.9 & 38.9(9.6 ) \\
0.6/0.6 & 19.2 & 6.2  & 2.9 & 28.4(10.6) \\
0.6/0.9 & 14.4 & 10.6 & 2.9 & 27.9(13.0) \\
0.8/0.3 & 47.1 & 3.4  & 2.4 & 52.9(3.4 ) \\
0.8/0.6 & 32.7 & 4.3  & 3.4 & 40.4(7.7 ) \\
0.8/0.9 & 21.6 & 6.7  & 4.3 & 32.7(5.8 ) \\
\hline
Average & 21.1 & 7.6  & 3.3 & 31.9(12.7) \\
\hline
\end{tabular}
\end{table}

\begin{table}
\begin{tabular}{lcccl}
\hline
$m_1 = 10^{10} \ \rm M_{\odot}$ & \multicolumn{4}{c}{\% Mergers} \\
$\iota$ 	 & $m_1$-$m_2$  & $m_1$-$m_3$  & $m_2$-$m_3$ & Total\\
\hline
     10 &  14.6 &  9.4 &  7.8 &   31.8(14.1) \\
34.9323 &  11.5 &  8.9 &  3.6 &   24.0(14.6) \\
49.0917 &  16.7 &  7.3 &  4.2 &   28.1(12.0) \\
60.6678 &  20.8 &  7.8 &  2.1 &   30.7(14.6) \\
71.0409 &  23.4 &  9.4 &  3.6 &   36.5(10.4) \\
80.7981 &  23.4 &  8.9 &  3.6 &   35.9(12.5) \\
90.2902 &  29.7 &  6.2 &  2.1 &   38.0(12.5) \\
99.7903 &  33.9 &  4.7 &  2.6 &   41.1(10.9) \\
109.574 &  27.1 &  6.2 &  1.0 &   34.4(12.0) \\
    120 &  25.5 &  5.7 &  1.6 &   32.8(15.1) \\
131.681 &  13.5 &  9.9 &  2.1 &   25.5(9.9 ) \\
146.094 &  15.1 &  5.7 &  3.6 &   24.5(12.0) \\
174.231 &  18.8 &  8.3 &  4.7 &   31.8(14.1) \\
\hline
Average &  21.1 &  7.6 &  3.3 &   31.9(12.7) \\
\hline
\end{tabular}
\end{table}

%%%%%%%%%%%%%%%%%%%%%%%%%%%%%%%%%%%%%%%%%%%%%%%%%%%%%%%%%%%%%%%%%%%%%
%%%%%%%%%%%%%%%%%%%%%%%%%%%%%%%%%%%%%%%%%%%%%%%%%%%%%%%%%%%%%%%%%%%%%
%%%%%%%%%%%%%%%%%%%%%%%%%%%%%%%%%%%%%%%%%%%%%%%%%%%%%%%%%%%%%%%%%%%%%
%m1 = 1e9

\begin{table}
\caption{Results $m_1=10^{9}\rm M_{\odot}$}
\label{tab:m1_9}
\begin{tabular}{lcccl}
\hline
$m_1 = 10^{9} \ \rm M_{\odot}$ & \multicolumn{4}{c}{\% Mergers} \\
$q_{\rm in}/q_{\rm out}$	 & $m_1$-$m_2$  & $m_1$-$m_3$  & $m_2$-$m_3$ & Total \\
\hline
0.0316/0.0316 &   9.0 &   5.1 &   0.0 &   14.1(20.5) \\
0.0316/   0.1 &  23.7 &   3.2 &   0.0 &   26.9(13.5) \\
0.0316/0.3162 &   9.0 &   1.3 &   0.6 &   10.9(3.8 ) \\
0.0316/     1 &  25.0 &   0.0 &   1.9 &   26.9(0.0 ) \\
   0.1/0.0316 &   4.5 &   1.9 &   0.0 &    6.4(16.7) \\
   0.1/   0.1 &  16.0 &   5.8 &   0.6 &   22.4(17.3) \\
   0.1/0.3162 &  24.4 &   3.2 &   1.3 &   28.8(11.5) \\
   0.1/     1 &  25.6 &   0.0 &   7.1 &   32.7(1.3 ) \\
0.3162/0.0316 &   2.6 &   0.6 &   0.0 &    3.2(7.1 ) \\
0.3162/   0.1 &  12.8 &   2.6 &   0.6 &   16.0(14.1) \\
0.3162/0.3162 &  26.3 &  14.1 &   3.8 &   44.2(16.0) \\
0.3162/     1 &  17.3 &   1.9 &   4.5 &   23.7(5.8 ) \\
     1/0.0316 &   1.3 &   0.0 &   0.0 &    1.3(5.1 ) \\
     1/   0.1 &   5.8 &   0.0 &   0.6 &    6.4(8.3 ) \\
     1/0.3162 &  26.9 &  12.2 &  14.1 &   53.2(16.0) \\
     1/     1 &  13.5 &  14.1 &  15.4 &   42.9(21.8) \\
\hline
Average		  &  15.2 &   4.1 &   3.2 &   22.5(11.2) \\
\hline
\end{tabular}
\end{table}

\begin{table}
\begin{tabular}{lcccl}
\hline
$m_1 = 10^{9} \ \rm M_{\odot}$ & \multicolumn{4}{c}{\% Mergers} \\
$e_{\rm in}/e_{\rm out}$ 	 & $m_1$-$m_2$  & $m_1$-$m_3$  & $m_2$-$m_3$ & Total \\
\hline
0.2/0.3 & 18.3 & 3.8 & 3.4 & 25.5(9.1 ) \\
0.2/0.6 & 7.7  & 3.8 & 3.4 & 14.9(8.7 ) \\
0.2/0.9 & 5.3  & 5.3 & 2.4 & 13.0(2.4 ) \\
0.4/0.3 & 16.3 & 3.4 & 3.8 & 23.6(13.5) \\
0.4/0.6 & 13.0 & 3.8 & 2.4 & 19.2(12.0) \\
0.4/0.9 & 7.7  & 5.3 & 3.4 & 16.3(8.2 ) \\
0.6/0.3 & 22.6 & 2.4 & 6.2 & 31.2(6.2 ) \\
0.6/0.6 & 12.0 & 5.3 & 2.9 & 20.2(11.1) \\
0.6/0.9 & 9.1  & 3.8 & 1.0 & 13.9(13.0) \\
0.8/0.3 & 34.1 & 1.9 & 1.9 & 38.0(5.3 ) \\
0.8/0.6 & 24.5 & 4.3 & 4.3 & 33.2(7.2 ) \\
0.8/0.9 & 12.0 & 6.2 & 2.9 & 21.2(13.0) \\
\hline
Average & 15.2 & 4.1 & 3.2 & 22.5(11.2) \\
\hline
\end{tabular}
\end{table}

\begin{table}
\begin{tabular}{lcccl}
\hline
$m_1 = 10^{9} \ \rm M_{\odot}$ & \multicolumn{4}{c}{\% Mergers} \\
$\iota$ & $m_1$-$m_2$  & $m_1$-$m_3$  & $m_2$-$m_3$ & Total\\
\hline
     10 &  10.9 &  4.2 &  6.2 &   21.4(14.6) \\ 
34.9323 &  10.4 &  4.2 &  3.1 &   17.7(10.4) \\
49.0917 &  10.4 &  1.6 &  7.8 &   19.8(14.1) \\
60.6678 &  10.4 &  2.1 &  7.3 &   19.8(13.0) \\
71.0409 &   8.3 &  5.7 &  3.6 &   17.7(8.3 ) \\
80.7981 &  17.7 &  6.2 &  0.5 &   24.5(12.5) \\
90.2902 &  26.6 &  6.2 &  1.0 &   33.9(8.3 ) \\
99.7903 &  27.1 &  2.1 &  1.0 &   30.2(8.9 ) \\
109.574 &  18.2 &  4.7 &  0.5 &   23.4(12.5) \\
    120 &  22.4 &  2.6 &  3.1 &   28.1(13.0) \\
131.681 &  14.1 &  4.2 &  1.0 &   19.3(8.9 ) \\
146.094 &  10.4 &  5.2 &  2.1 &   17.7(10.9) \\
174.231 &  10.9 &  4.7 &  3.6 &   19.3(9.9 ) \\
\hline
Average &  15.2 &  4.1 &  3.2 &   22.5(11.2) \\
\hline
\end{tabular}
\end{table}

%%%%%%%%%%%%%%%%%%%%%%%%%%%%%%%%%%%%%%%%%%%%%%%%%%%%%%%%%%%%%%%%%%%%%
%%%%%%%%%%%%%%%%%%%%%%%%%%%%%%%%%%%%%%%%%%%%%%%%%%%%%%%%%%%%%%%%%%%%%
%%%%%%%%%%%%%%%%%%%%%%%%%%%%%%%%%%%%%%%%%%%%%%%%%%%%%%%%%%%%%%%%%%%%%
%m1 = 1e8

\begin{table}
\caption{Results $m_1=10^{8}\rm M_{\odot}$}
\label{tab:m1_8}
\begin{tabular}{lcccl}
\hline
$m_1 = 10^{8} \ \rm M_{\odot}$ & \multicolumn{4}{c}{\% Mergers} \\
$q_{\rm in}/q_{\rm out}$	 & $m_1$-$m_2$  & $m_1$-$m_3$  & $m_2$-$m_3$ & Total \\
\hline
0.0316/0.0316 &   9.6 &   8.3 &   0.0 &   17.9(5.1 ) \\ 
0.0316/   0.1 &  14.1 &   2.6 &   0.0 &   16.7(8.3 ) \\
0.0316/0.3162 &  28.2 &   0.6 &   0.6 &   29.5(1.9 ) \\
0.0316/     1 &  14.7 &   1.3 &   2.6 &   18.6(3.2 ) \\
   0.1/0.0316 &   4.5 &   3.2 &   0.0 &    7.7(5.8 ) \\
   0.1/   0.1 &   9.6 &   6.4 &   0.0 &   16.0(12.8) \\
   0.1/0.3162 &  22.4 &   6.4 &   1.3 &   30.1(5.1 ) \\
   0.1/     1 &  21.8 &   0.0 &   1.9 &   23.7(2.6 ) \\
0.3162/0.0316 &   1.3 &   1.3 &   0.0 &    2.6(3.2 ) \\
0.3162/   0.1 &  10.3 &   1.3 &   0.6 &   12.2(7.1 ) \\
0.3162/0.3162 &  25.6 &   9.0 &   1.9 &   36.5(11.5) \\
0.3162/     1 &  23.1 &   3.2 &   8.3 &   34.6(5.1 ) \\
     1/0.0316 &   0.0 &   0.6 &   0.0 &    0.6(3.2 ) \\
     1/   0.1 &   5.8 &   0.0 &   0.0 &    5.8(5.1 ) \\
     1/0.3162 &  25.0 &   9.6 &  10.9 &   45.5(11.5) \\
     1/     1 &  19.2 &  10.3 &  11.5 &   41.0(9.0 ) \\
\hline
Average		  &  14.7 &   4.0 &   2.5 &   21.2(6.3 ) \\
\hline
\end{tabular}
\end{table}

\begin{table}
\begin{tabular}{lcccl}
\hline
$m_1 = 10^{8} \ \rm M_{\odot}$ & \multicolumn{4}{c}{\% Mergers} \\
$e_{\rm in}/e_{\rm out}$ 	 & $m_1$-$m_2$  & $m_1$-$m_3$  & $m_2$-$m_3$ & Total \\
\hline
0.2/0.3 & 15.9 & 4.3 & 4.8 & 25.0(3.4 ) \\
0.2/0.6 & 12.0 & 4.3 & 3.4 & 19.7(6.2 ) \\
0.2/0.9 & 4.3  & 3.8 & 2.4 & 10.6(2.4 ) \\
0.4/0.3 & 18.3 & 3.4 & 1.9 & 23.6(5.3 ) \\
0.4/0.6 & 14.9 & 4.8 & 1.4 & 21.2(11.1) \\
0.4/0.9 & 4.8  & 5.8 & 2.9 & 13.5(5.3 ) \\
0.6/0.3 & 20.7 & 2.9 & 1.4 & 25.0(1.4 ) \\
0.6/0.6 & 13.0 & 3.8 & 1.4 & 18.3(4.3 ) \\
0.6/0.9 & 3.4  & 5.3 & 2.9 & 11.5(1.4 ) \\
0.8/0.3 & 34.6 & 1.4 & 2.4 & 38.5(1.0 ) \\
0.8/0.6 & 23.6 & 4.3 & 1.9 & 29.8(2.4 ) \\
0.8/0.9 & 11.1 & 3.8 & 2.9 & 17.8(5.8 ) \\
\hline
Average & 14.7 & 4.0 & 2.5 & 21.2(6.3 ) \\
\hline
\end{tabular}
\end{table}

\begin{table}
\begin{tabular}{lcccl}
\hline
$m_1 = 10^{8} \ \rm M_{\odot}$ & \multicolumn{4}{c}{\% Mergers} \\
$\iota$ & $m_1$-$m_2$  & $m_1$-$m_3$  & $m_2$-$m_3$ & Total\\
\hline
     10 &  10.4 &  3.1 &  3.6 &   17.2(6.8) \\
34.9323 &   8.9 &  3.1 &  3.1 &   15.1(6.2) \\
49.0917 &   6.8 &  5.2 &  4.2 &   16.1(8.3) \\
60.6678 &  13.0 &  5.7 &  2.6 &   21.4(4.7) \\
71.0409 &  11.5 &  3.1 &  2.6 &   17.2(8.9) \\
80.7981 &  14.6 &  2.6 &  4.2 &   21.4(6.8) \\
90.2902 &  27.6 &  0.5 &  1.6 &   29.7(3.1) \\
99.7903 &  28.1 &  2.1 &  1.0 &   31.2(3.1) \\
109.574 &  20.3 &  4.2 &  2.1 &   26.6(6.8) \\
    120 &  19.3 &  2.6 &  1.6 &   23.4(5.2) \\
131.681 &  13.5 &  4.2 &  2.1 &   19.8(5.2) \\
146.094 &   7.8 &  7.3 &  0.0 &   15.1(8.9) \\
174.231 &   9.4 &  8.3 &  3.6 &   21.4(7.8) \\
\hline
Average &  14.7 &  4.0 &  2.5 &   21.2(6.3) \\
\hline
\end{tabular}
\end{table}

%%%%%%%%%%%%%%%%%%%%%%%%%%%%%%%%%%%%%%%%%%%%%%%%%%%%%%%%%%%%%%%%%%%%%
%%%%%%%%%%%%%%%%%%%%%%%%%%%%%%%%%%%%%%%%%%%%%%%%%%%%%%%%%%%%%%%%%%%%%
%%%%%%%%%%%%%%%%%%%%%%%%%%%%%%%%%%%%%%%%%%%%%%%%%%%%%%%%%%%%%%%%%%%%%
%m1 = 10^7

\begin{table}
\caption{Results $m_1=10^{7}\rm M_{\odot}$}
\label{tab:m1_7}
\begin{tabular}{lcccl}
\hline
$m_1 = 10^{7} \ \rm M_{\odot}$ & \multicolumn{4}{c}{\% Mergers} \\
$q_{\rm in}/q_{\rm out}$	 & $m_1$-$m_2$  & $m_1$-$m_3$  & $m_2$-$m_3$ & Total \\
\hline
0.0316/0.0316 &   5.1 &  4.5 &  0.0 &    9.6(6.4 ) \\
0.0316/   0.1 &  23.1 &  1.9 &  0.0 &   25.0(1.9 ) \\
0.0316/0.3162 &  23.7 &  3.2 &  0.6 &   27.6(7.7 ) \\
0.0316/     1 &  23.7 &  5.8 &  0.0 &   29.5(1.9 ) \\
   0.1/0.0316 &   3.2 &  0.6 &  0.0 &    3.8(3.2 ) \\
   0.1/   0.1 &   9.6 &  2.6 &  0.0 &   12.2(6.4 ) \\
   0.1/0.3162 &  32.1 &  1.3 &  0.6 &   34.0(5.1 ) \\
   0.1/     1 &  22.4 &  1.9 &  0.6 &   25.0(3.2 ) \\
0.3162/0.0316 &   0.0 &  0.6 &  0.0 &    0.6(0.6 ) \\
0.3162/   0.1 &  10.3 &  0.0 &  1.3 &   11.5(5.8 ) \\
0.3162/0.3162 &  23.1 &  6.4 &  0.6 &   30.1(4.5 ) \\
0.3162/     1 &  23.1 &  0.0 &  7.1 &   30.1(0.6 ) \\
     1/0.0316 &   1.3 &  0.0 &  0.0 &    1.3(0.6 ) \\
     1/   0.1 &   6.4 &  1.3 &  0.6 &    8.3(2.6 ) \\
     1/0.3162 &  24.4 &  3.8 &  5.1 &   33.3(9.0 ) \\
     1/     1 &  14.7 &  6.4 &  6.4 &   27.6(11.5) \\
\hline
Average		  &  15.4 &  2.5 &  1.4 &   19.4(4.4 ) \\
\hline
\end{tabular}
\end{table}

\begin{table}
\begin{tabular}{lcccl}
\hline
$m_1 = 10^{7} \ \rm M_{\odot}$ & \multicolumn{4}{c}{\% Mergers} \\
$e_{\rm in}/e_{\rm out}$  	 & $m_1$-$m_2$  & $m_1$-$m_3$  & $m_2$-$m_3$ & Total \\
\hline
0.2/0.3 & 18.8 & 2.4 & 2.4 & 23.6(2.4) \\
0.2/0.6 & 6.7  & 2.4 & 1.9 & 11.1(0.0) \\
0.2/0.9 & 3.4  & 4.8 & 1.4 & 9.6 (1.4) \\
0.4/0.3 & 21.6 & 1.4 & 2.4 & 25.5(3.8) \\
0.4/0.6 & 13.9 & 1.0 & 1.0 & 15.9(6.2) \\
0.4/0.9 & 5.8  & 2.9 & 1.4 & 10.1(4.8) \\
0.6/0.3 & 21.6 & 2.4 & 1.0 & 25.0(1.4) \\
0.6/0.6 & 13.5 & 1.9 & 1.9 & 17.3(5.8) \\
0.6/0.9 & 8.7  & 2.9 & 0.5 & 12.0(3.8) \\
0.8/0.3 & 39.4 & 0.5 & 0.5 & 40.4(1.4) \\
0.8/0.6 & 22.6 & 4.8 & 1.4 & 28.8(1.9) \\
0.8/0.9 & 8.7  & 2.9 & 1.4 & 13.0(4.3) \\
\hline
Average & 15.4 & 2.5 & 1.4 & 19.4(4.4) \\
\hline
\end{tabular}
\end{table}

\begin{table}
\begin{tabular}{lcccl}
\hline
$m_1 = 10^{7} \ \rm M_{\odot}$ & \multicolumn{4}{c}{\% Mergers} \\
$\iota$ 	 & $m_1$-$m_2$  & $m_1$-$m_3$  & $m_2$-$m_3$ & Total\\
\hline
     10 &  12.5 &  2.6 &  1.6 &   16.7(5.7) \\
34.9323 &  10.4 &  1.0 &  2.1 &   13.5(6.8) \\
49.0917 &  12.0 &  2.1 &  3.1 &   17.2(3.6) \\
60.6678 &  11.5 &  3.1 &  1.6 &   16.1(4.7) \\
71.0409 &  14.6 &  2.6 &  1.6 &   18.8(5.2) \\
80.7981 &  14.1 &  4.2 &  1.0 &   19.3(4.7) \\
90.2902 &  31.2 &  1.6 &  1.0 &   33.9(3.6) \\
99.7903 &  26.0 &  1.0 &  1.0 &   28.1(3.1) \\
109.574 &  24.0 &  3.6 &  0.0 &   27.6(2.6) \\
    120 &  17.2 &  1.6 &  1.0 &   19.8(4.7) \\
131.681 &  11.5 &  2.1 &  1.6 &   15.1(4.2) \\
146.094 &   8.3 &  3.6 &  1.0 &   13.0(4.2) \\
174.231 &   6.8 &  3.6 &  2.1 &   12.5(4.7) \\
\hline
Average &  15.4 &  2.5 &  1.4 &   19.4(4.4) \\
\hline
\end{tabular}
\end{table}

%%%%%%%%%%%%%%%%%%%%%%%%%%%%%%%%%%%%%%%%%%%%%%%%%%%%%%%%%%%%%%%%%%%%%
%%%%%%%%%%%%%%%%%%%%%%%%%%%%%%%%%%%%%%%%%%%%%%%%%%%%%%%%%%%%%%%%%%%%%
%%%%%%%%%%%%%%%%%%%%%%%%%%%%%%%%%%%%%%%%%%%%%%%%%%%%%%%%%%%%%%%%%%%%%
%m1 = 1e6

\begin{table}
\caption{Results $m_1=10^{6}\rm M_{\odot}$}
\label{tab:m1_6}
\begin{tabular}{lcccl}
\hline
$m_1 = 10^{6} \ \rm M_{\odot}$ & \multicolumn{4}{c}{\% Mergers} \\
$q_{\rm in}/q_{\rm out}$	 & $m_1$-$m_2$  & $m_1$-$m_3$  & $m_2$-$m_3$ & Total \\
\hline
0.0316/0.0316 &   4.5 &  0.6 &  0.0 &    5.1(1.3) \\
0.0316/   0.1 &  11.5 &  0.0 &  0.0 &   11.5(0.0) \\
0.0316/0.3162 &  25.6 &  0.0 &  0.0 &   25.6(2.6) \\
0.0316/     1 &  39.7 &  0.0 &  1.9 &   41.7(0.6) \\
   0.1/0.0316 &   1.3 &  0.6 &  0.0 &    1.9(1.9) \\
   0.1/   0.1 &   9.0 &  1.3 &  0.0 &   10.3(6.4) \\
   0.1/0.3162 &  38.5 &  0.0 &  0.0 &   38.5(2.6) \\
   0.1/     1 &  43.6 &  0.0 &  0.6 &   44.2(0.6) \\
0.3162/0.0316 &   0.6 &  0.6 &  0.0 &    1.3(1.3) \\
0.3162/   0.1 &   3.2 &  0.6 &  0.0 &    3.8(0.0) \\
0.3162/0.3162 &  19.9 &  1.3 &  0.0 &   21.2(1.9) \\
0.3162/     1 &  26.3 &  1.3 &  2.6 &   30.1(1.9) \\
     1/0.0316 &   0.6 &  0.0 &  0.0 &    0.6(0.6) \\
     1/   0.1 &   1.9 &  0.6 &  0.6 &    3.2(1.3) \\
     1/0.3162 &  17.9 &  6.4 &  4.5 &   28.8(5.1) \\
     1/     1 &  14.7 &  8.3 &  5.8 &   28.8(1.9) \\
\hline
Average		  &  16.2 &  1.4 &  1.0 &   18.5(1.9) \\
\hline
\end{tabular}
\end{table}

\begin{table}
\begin{tabular}{lcccl}
\hline
$m_1 = 10^{6} \ \rm M_{\odot}$ & \multicolumn{4}{c}{\% Mergers} \\
$e_{\rm in}/e_{\rm out}$  	 & $m_1$-$m_2$  & $m_1$-$m_3$  & $m_2$-$m_3$ & Total \\
\hline
0.2/0.3 &  16.3 & 2.9 & 1.4 & 20.7(1.4) \\
0.2/0.6 &  6.7  & 0.5 & 0.5 & 7.7 (1.4) \\
0.2/0.9 &  6.2  & 1.0 & 1.4 & 8.7 (3.4) \\
0.4/0.3 &  18.8 & 1.4 & 1.4 & 21.6(1.9) \\
0.4/0.6 &  14.4 & 1.4 & 1.9 & 17.8(3.4) \\
0.4/0.9 &  7.7  & 1.0 & 1.0 & 9.6 (3.4) \\
0.6/0.3 &  27.4 & 0.5 & 0.5 & 28.4(0.0) \\
0.6/0.6 &  13.9 & 2.4 & 0.0 & 16.3(0.0) \\
0.6/0.9 &  9.1  & 1.4 & 0.5 & 11.1(0.0) \\
0.8/0.3 &  38.0 & 1.4 & 1.4 & 40.9(0.0) \\
0.8/0.6 &  22.6 & 0.5 & 0.5 & 23.6(0.0) \\
0.8/0.9 &  13.0 & 1.9 & 1.4 & 16.3(0.5) \\
\hline
Average &  16.2 & 1.4 & 1.0 & 18.5(1.9) \\
\hline
\end{tabular}
\end{table}

\begin{table}
\begin{tabular}{lcccl}
\hline
$m_1 = 10^{6} \ \rm M_{\odot}$ & \multicolumn{4}{c}{\% Mergers} \\
$\iota$ & $m_1$-$m_2$  & $m_1$-$m_3$  & $m_2$-$m_3$ & Total\\
\hline
     10 &  13.0 &  3.1 &  1.0 &   17.2(3.6) \\
34.9323 &  13.5 &  0.0 &  2.1 &   15.6(1.6) \\
49.0917 &  15.6 &  2.1 &  1.0 &   18.8(1.0) \\
60.6678 &  17.7 &  0.5 &  1.6 &   19.8(2.6) \\
71.0409 &  14.1 &  0.5 &  1.0 &   15.6(1.6) \\
80.7981 &  18.2 &  0.5 &  0.0 &   18.8(1.6) \\
90.2902 &  33.3 &  0.5 &  0.5 &   34.4(0.5) \\
99.7903 &  25.5 &  0.5 &  0.0 &   26.0(1.6) \\
109.574 &  20.3 &  1.0 &  0.5 &   21.9(2.1) \\
    120 &  15.6 &  2.6 &  1.6 &   19.8(0.5) \\
131.681 &   9.9 &  1.6 &  0.0 &   11.5(4.2) \\
146.094 &   5.7 &  1.6 &  3.1 &   10.4(1.0) \\
174.231 &   7.8 &  3.1 &  0.5 &   11.5(2.6) \\
\hline
Average &  16.2 &  1.4 &  1.0 &   18.5(1.9) \\
\hline
\end{tabular}
\end{table}

%%%%%%%%%%%%%%%%%%%%%%%%%%%%%%%%%%%%%%%%%%%%%%%%%%%%%%%%%%%%%%%%%%%%%
%%%%%%%%%%%%%%%%%%%%%%%%%%%%%%%%%%%%%%%%%%%%%%%%%%%%%%%%%%%%%%%%%%%%%
%%%%%%%%%%%%%%%%%%%%%%%%%%%%%%%%%%%%%%%%%%%%%%%%%%%%%%%%%%%%%%%%%%%%%
%m1 = 1e5

\begin{table}
\centering
\caption{Results $m_1=10^{5}\rm M_{\odot}$}
\label{tab:m1_5}
\begin{tabular}{lcccl}
\hline
$m_1 = 10^{5} \ \rm M_{\odot}$ & \multicolumn{4}{c}{\% Mergers} \\
$q_{\rm in}/q_{\rm out}$	 & $m_1$-$m_2$  & $m_1$-$m_3$  & $m_2$-$m_3$ & Total \\
\hline
0.0316/0.0316 &   1.9 &  1.9 &  0.0 &    3.8(2.6) \\
0.0316/   0.1 &  19.2 &  1.3 &  0.0 &   20.5(0.0) \\
0.0316/0.3162 &  34.6 &  0.0 &  0.0 &   34.6(1.3) \\
0.0316/     1 &  23.1 &  0.0 &  1.3 &   24.4(3.8) \\
   0.1/0.0316 &   0.6 &  0.0 &  0.0 &    0.6(1.3) \\
   0.1/   0.1 &   8.3 &  0.6 &  0.6 &    9.6(3.2) \\
   0.1/0.3162 &  35.3 &  0.6 &  0.0 &   35.9(1.3) \\
   0.1/     1 &  40.4 &  0.0 &  0.0 &   40.4(0.0) \\
0.3162/0.0316 &   1.9 &  0.0 &  0.0 &    1.9(0.0) \\
0.3162/   0.1 &   2.6 &  0.6 &  0.0 &    3.2(3.8) \\
0.3162/0.3162 &  28.2 &  0.6 &  0.6 &   29.5(3.2) \\
0.3162/     1 &  30.8 &  0.6 &  3.8 &   35.3(0.0) \\
     1/0.0316 &   0.0 &  0.0 &  0.0 &    0.0(0.6) \\
     1/   0.1 &   5.8 &  0.0 &  0.0 &    5.8(1.3) \\
     1/0.3162 &  19.2 &  3.8 &  5.1 &   28.2(2.6) \\
     1/     1 &  16.7 &  4.5 &  1.9 &   23.1(1.3) \\
\hline
Average		  &  16.8 &  0.9 &  0.8 &   18.5(1.6) \\
\hline
\end{tabular}
\end{table}

\begin{table}
\begin{tabular}{lcccl}
\hline
$m_1 = 10^{5} \ \rm M_{\odot}$ & \multicolumn{4}{c}{\% Mergers} \\
$e_{\rm in}/e_{\rm out}$  	 & $m_1$-$m_2$  & $m_1$-$m_3$  & $m_2$-$m_3$ & Total \\
\hline
0.2/0.3 &  17.3 &  0.5 &  1.4 & 19.2(1.0) \\
0.2/0.6 &  10.1 &  0.5 &  0.5 & 11.1(0.5) \\
0.2/0.9 &  2.9  &  1.9 &  1.0 & 5.8 (1.4) \\
0.4/0.3 &  17.8 &  0.0 &  1.9 & 19.7(0.5) \\
0.4/0.6 &  12.5 &  1.4 &  1.0 & 14.9(1.4) \\
0.4/0.9 &  6.7  &  1.0 &  0.5 & 8.2 (1.4) \\
0.6/0.3 &  26.4 &  1.4 &  1.0 & 28.8(0.5) \\
0.6/0.6 &  18.8 &  0.5 &  0.5 & 19.7(1.0) \\
0.6/0.9 &  10.1 &  0.5 &  0.5 & 11.1(2.4) \\
0.8/0.3 &  38.5 &  0.0 &  0.5 & 38.9(1.0) \\
0.8/0.6 &  26.0 &  1.0 &  1.0 & 27.9(1.4) \\
0.8/0.9 &  14.4 &  2.4 &  0.5 & 17.3(2.4) \\
\hline
Average &  16.8 &  0.9 &  0.8 & 18.5(1.6) \\
\hline
\end{tabular}
\end{table}

\begin{table}
\begin{tabular}{lcccl}
\hline
$m_1 = 10^{5} \ \rm M_{\odot}$ & \multicolumn{4}{c}{\% Mergers} \\
$\iota$ 	 & $m_1$-$m_2$  & $m_1$-$m_3$  & $m_2$-$m_3$ & Total\\
\hline
     10 &  16.7 &  0.0 &  1.6 &   18.2(3.1) \\
34.9323 &  15.6 &  1.6 &  0.5 &   17.7(2.6) \\
49.0917 &  18.2 &  0.0 &  1.6 &   19.8(1.6) \\
60.6678 &  15.1 &  0.5 &  2.6 &   18.2(1.6) \\
71.0409 &  12.5 &  0.5 &  0.5 &   13.5(2.6) \\
80.7981 &  16.1 &  0.0 &  0.5 &   16.7(1.0) \\
90.2902 &  31.8 &  0.5 &  0.5 &   32.8(1.6) \\
99.7903 &  30.7 &  0.5 &  0.0 &   31.2(1.0) \\
109.574 &  21.4 &  1.6 &  2.1 &   25.0(1.0) \\
    120 &  17.7 &  0.5 &  0.0 &   18.2(1.0) \\
131.681 &   9.9 &  1.6 &  0.5 &   12.0(0.5) \\
146.094 &   5.2 &  1.6 &  0.0 &    6.8(2.1) \\
174.231 &   7.3 &  3.1 &  0.5 &   10.9(1.6) \\
\hline
Average &  16.8 &  0.9 &  0.8 &   18.5(1.6) \\
\hline
\end{tabular}
\end{table}
%%%%%%%%%%%%%%%%%%%%%%%%%%%%%%%%%%%%%%%%%%%%%%

\end{document}